\documentstyle[preprint,aps,epsf,floats]{revtex}

\newcommand{\nn}{\nonumber}
\newcommand{\ppp}{\mbox{$({\mathbf p'}-{\mathbf p})^2$}}

\newcommand{\ply}{ {\rm Li}_2 }

\def\xslash#1{{\rlap{$#1$}/}}
\def\bsigma{\mbox{\boldmath $\sigma$}}

\def\psip#1{\psi_{\mathbf{#1}}}
\def\chip#1{\chi_{\mathbf{#1}}}
\def\abs#1{\left| #1 \right|}
\def\OMIT#1{}

\begin{document}
\setlength\baselineskip{20pt}

\preprint{\tighten  \hbox{UCSD/PTH 00-05}
}

\title{Running of the heavy quark production current and\\
$1/|{\bf k}|$ potential in QCD}

\author{Aneesh V. Manohar\footnote{amanohar@ucsd.edu} and
Iain W.\ Stewart\footnote{iain@schwinger.ucsd.edu} \\[4pt]}
\address{\tighten Department of Physics, University of California at San
Diego,\\[2pt] 9500 Gilman Drive, La Jolla, CA 92093 }

\maketitle

{\tighten
\begin{abstract}
The $1/|{\bf k}|$ contribution to the heavy quark potential is first generated
at one loop order in QCD. We compute the two loop anomalous dimension for this
potential, and find that the renormalization group running is significant. The
next-to-leading-log coefficient for the heavy quark production current near
threshold is determined. The velocity renormalization group result reproduces
the $\alpha_s^3 \ln^2(\alpha_s)$ ``non-renormalization group logarithms'' of
Kniehl and Penin.
\end{abstract}

}
\vspace{0.7in}

\newpage 

\section{Introduction}

For systems involving a heavy quark-antiquark pair near threshold it is useful
to combine the QCD coupling constant expansion with an expansion in powers of
the relative velocity $v$.  This expansion is facilitated by using
non-relativistic QCD formulated as an effective field theory with an explicit
power counting in
$v$~\cite{Caswell,BBL,Labelle,lm,Manohar,gr,ls,Pineda,P2,Beneke,Gries,P3,kp,LMR}.
For the potential between a heavy quark and antiquark this double expansion
takes the form
\begin{eqnarray} \label{V1}
  & & V({\bf p},{\bf p'}) = \sum_{n=-2}^{\infty} V^{(n)} , \qquad
     V^{(n)} = \sum_{j=1}^\infty  V^{(n,j)} , \nn\\[5pt]
  & & \mbox{where}\quad V^{(n)}\sim  v^n \,, \qquad\quad
     V^{(n,j)}\sim  v^n \alpha_s^j  \,.
\end{eqnarray}
The even terms $V^{(2k)}$ are first generated at tree-level (order $\alpha_s$),
and the odd terms $V^{(2k+1)}$ are first generated at one loop (order
$\alpha_s^2$).

Matrix elements for non-relativistic QCD systems typically depend on logarithms
of the velocity $v$. For small $v$ it is convenient to sum large logarithms
of the form $\alpha_s(mv) \ln(v)$ and $\alpha_s(mv^2) \ln(v)$ by using
renormalization group equations in the effective theory.  This reorganizes the
series in $j$ in Eq.~(\ref{V1}) so that:
\begin{eqnarray} \label{VLL}
  V^{(n)} = \sum_{j} \tilde V^{(n,j)}\,,\qquad\quad \mbox{where} \qquad
   \tilde V^{(n,j)} \sim v^n\, \alpha_s(m)^j\ \sum_{k=0}^\infty \:
   [\alpha_s(m) \ln(v)]^k \:  \,.
\end{eqnarray}
The simplest example of such a summation is the use of a running coupling
constant, $\alpha_s(\mu)$, instead of the coupling at the matching scale
$\alpha_s(m)$. For $\mu<m$ the running coupling includes a series of
$\alpha_s(m) \ln(\mu/m)$ terms. However, it should be emphasized that the
complete set of renormalization group logarithms are not determined by the
simple replacement $\alpha_s(m) \to \alpha_s(\mu)$.  

A complication in summing the logarithms is the presence of two low energy
scales: the typical momenta of the heavy quark $\sim m v$ (soft scale), and
typical energy $\sim m v^2$ (ultrasoft scale). Two approaches have been proposed
for dealing with the presence of two scales, one stage and two stage
running. With two stage running, one matches QCD onto an effective theory at the
scale $\mu=m$ and then runs to the scale $\mu=m v$. At $\mu=m v$ one matches
onto an effective theory called pNRQCD~\cite{Pineda,Brambilla} which has
composite $q\bar q$ fields, and then considers the running to $\mu=m v^2$. In
vNRQCD~\cite{LMR}, the running occurs in a single stage using a velocity
renormalization group. The velocity renormalization group takes into account
that the scales $mv$ and $mv^2$ are tied together by the heavy quark equation of
motion for all $\mu<m$. We use dimensional regularization and the $\overline{\rm
MS}$ scheme with both the ultrasoft scale $\mu_U\sim mv^2$ and the soft scale
$\mu_S\sim mv$ given in terms of a single subtraction point velocity $\nu$:
$\mu_U=m\nu^2$ and $\mu_S=m\nu$. The renormalization group equations are written
for the variable $\nu$. It is an interesting question as to whether both the one
and two stage methods of running will sum the full set of $\alpha_s \ln(v)$
terms. In this paper only the single stage running will be considered.

The running of the Coulomb potential at one and two loops, $\tilde V^{(-2,1)}$
and $\tilde V^{(-2,2)}$, is determined by the running coupling constant
$\alpha_s(\mu)~$\cite{Peter} \footnote{\tighten At three loops
ultrasoft gluons can contribute to the running of the static
potential\cite{psIR}.}. The summation of the leading logarithms for the the
$v^0$ potential was carried out in Refs.~\cite{chen,amis}. In this paper we
extend this analysis to the $1/|{\bf k}|$ potential by calculating its two loop
anomalous dimension.

If $m v^2 \gg \Lambda_{\rm QCD}$, the $v$ expansion can be applied to
non-relativistic QCD systems in a perturbative manner. This is the
case for $t\bar t$ production near threshold where $m v^2\sim 4\,{\rm GeV}$,
and is the situation that will be analyzed in this paper. Of particular
interest is the Coulombic regime where $v\sim \alpha_s$.  In this regime the
expansion in Eq.~(\ref{V1}) has the form
\begin{eqnarray} \label{Vbnd}
 V &=& \Big[ \tilde V^{(-2,1)} \Big] + \Big[\tilde V^{(-2,2)} \Big] +
    \Big[ \tilde V^{(-2,3)} +\tilde V^{(-1,2)} +\tilde V^{(0,1)} \Big] +
    \ldots  \nn \\
  & \sim & \bigg[\: {\alpha_s \over v^2} \: \bigg] \ +\
  \bigg[\: { \alpha_s^2 \over v^2} \: \bigg] \ + \
  \bigg[\: { \alpha_s^3 \over v^2} + {\alpha_s^2 \over v}
  + {\alpha_s v^0 } \: \bigg] \ +  \ldots \,.
\end{eqnarray}
If the $\alpha_s \ln(v)$ dependence is treated perturbatively these terms are
referred to as the leading order (LO), next-to-leading order (NLO), and
next-to-next-to-leading order (NNLO) potentials. Note that since the $1/|{\bf
k}|$ potential first occurs at one loop, it only contributes at NNLO. When the
series in $\alpha_s \ln v$ are summed, the terms in Eq.~(\ref{Vbnd}) will be
referred to as leading-log (LL), next-to-leading log (NLL) and
next-to-next-to-leading log (NNLL), respectively. In the Coulomb regime, the
Coulomb potential must be kept to all orders.  Each additional Coulomb insertion
gives a $\alpha_s/v^2$ plus a factor of $v$ (from the potential loop), so each
new Coulomb interaction costs a factor of $\alpha_s/v\sim 1$.

To study the threshold production of $t\bar t$, a non-relativistic expansion
must also be made for the electromagnetic production current\footnote{\tighten
We will ignore effects associated with the top quark width.}
\begin{eqnarray} \label{pcurrent}
  \bar t\: \gamma^i\: t &=& \sum_{\bf p} c_1\: \Big( \psi^\dagger_{\bf p}
  \bsigma^i \chi^*_{\bf -p} \Big)  - \frac{c_2}{2 m^2} \Big( \psi^\dagger_{\bf p}
  \, {\bf p\cdot \bsigma \: p^i}\, \chi^*_{\bf -p} \Big)
   - \frac{c_3}{m^2} \Big( \psi^\dagger_{\bf p}\, {\bf p^2 \bsigma^i }\,
     \chi^*_{\bf -p} \Big) + \ldots \,.
 \end{eqnarray}
The fields $\psi^*$ and $\chi^*$ create non-relativistic top quarks and
antiquarks respectively. The $c_1$ term contributes at order $v^0$, and the
$c_2$ and $c_3$ terms contribute at order $v^2$. The current on the LHS of
Eq.~(\ref{pcurrent}) is conserved, and has no anomalous dimension in QCD.
However the non-relativistic current operators on the RHS are scale dependent
in the effective theory, and the coefficients $c_j$ therefore depend on
logarithms of $\mu$. The coefficients $c_j$ each have an expansion in
$\alpha_s$. The matching at $\mu\sim m$ is known to order $\alpha_s^2$ for
$c_1$\cite{pcQED,pcQCD}, and to order $\alpha_s$ for $c_2$ and $c_3$\cite{ls},
so the production current is known to NNLO with partial N$^3$LO results. At LL
order, one needs the tree-level matching $c_1=1$ at $\mu=m$, and the $v^2$
coefficients $c_2$ and $c_3$ can be set to zero. There is no one loop anomalous
dimension for the operator $\psi^\dagger_{\bf p} \sigma^i \chi^*_{\bf -p}$, so
the LL result is that $c_1=1$ at all $\nu$.

At NLL order, we need the one loop matching for $c_1$ at $\mu=m$, and the
two loop running for $c_1$. The coefficients $c_{2,3}$ first enter at NNLL, at
which order one would also need the three-loop anomalous dimension for $c_1$.
At two loops, the anomalous dimension for $c_1$ was computed at the matching
scale $\mu=m$\cite{pcQED,pcQCD}
\begin{eqnarray} \label{c1adm}
  \mu {\partial \over\partial\mu }\: c_1(\mu)\ \bigg|_{\mu=m} =
     - C_F \bigg( \frac13 C_F +\frac12 C_A \bigg) \alpha_s^2(m) \,,
\end{eqnarray}
by studying the two loop matching condition for $c_1$. For $\mu < m$ the
anomalous dimension no longer has the simple form Eq.~(\ref{c1adm}), but depends
on the running of the quark potential. This anomalous dimension was computed in
Ref.~\cite{LMR}, and depends on the running values $[\tilde V^{(-2,1)}]^2$,\,
${\tilde V^{(-2,1)}}\times {\tilde V^{(0,1)}}$, and $\tilde V^{(-1,2)}$ (see
Eq.~(\ref{c1ad}) below). It is interesting that the structure of this result
implies that determining the RHS of Eq.~(\ref{c1adm}) for $\mu<m$ requires the
LL values of $\tilde V^{(-2,1)}$ and $\tilde V^{(0,1)}$, but the NLL value of
$\tilde V^{(-1,2)}$. The one loop running of $\tilde V^{(-2,1)}$ is well known,
and the running of $\tilde V^{(-2,2)}$ was computed in Refs.~\cite{chen,amis}.
The two loop running of $\tilde V^{(-1,2)}$ is computed in this paper. Using the
running of these terms in the potential, we arrive at a complete NLL expression
for $c_1$.  The running of the non-relativistic scalar current is also briefly
discussed.

In section~II the effective theory is reviewed. We explain how
reparameterization invariance fixes the value of the lowest order coupling of an
ultrasoft gluon to the Coulomb potential.  The details of the computation of the
two loop anomalous dimension for the $1/|{\bf k}|$ potential are given in
section~III, and in the appendices. In section~III we give a derivation of the
anomalous dimension using on-shell potentials, while in appendix~\ref{App_off}
we repeat the derivation in the presence of off-shell potentials.  Readers not
interested in the technical details can skip to section~IV, where our results
are discussed.  In section~IV we expand our renormalization group improved
results in powers of $\alpha_s$ to compare to finite order calculations in the
literature.  For the color singlet $1/|{\bf k}|$ potential the first $\alpha_s
\ln(v)$ term in the series was computed in Ref.~\cite{Brambilla}, and our result
for this term agrees with theirs.  In Ref.~\cite{Penin}, Kniehl and Penin
computed the $\alpha_s^3 \ln^2(\alpha_s)$ terms in the wavefunction at the
origin which they refer to as ``non-renormalization group logarithms'', since
they do not involve factors of the $\beta$-function for $\alpha_s$. We show that
the second term in the series generated by our NLL production current agrees
with the result in Ref.~\cite{Penin}.  Thus, the solution of the renormalization
group equations in the velocity renormalization group method does include these
logarithms.  Finally, we discuss our result for the NLL $1/|{\bf k}|$ potential
and production current.

\section{The \lowercase{v}NRQCD Lagrangian}

The vNRQCD effective Lagrangian has the form \cite{LMR}
\begin{eqnarray}
 {\cal L} = {\cal L}_u + {\cal L}_p + {\cal L}_s \,.
\end{eqnarray}
The ultrasoft Lagrangian ${\cal L}_u$ involves the fields $\psip p$ which
annihilate a quark, $\chip p$ which annihilate an antiquark, and $A^\mu$ which
annihilate and create ultrasoft gluons.  The potential Lagrangian ${\cal L}_p$
contains operators with four or more quark fields including the quark-antiquark
potential. Finally, the soft Lagrangian ${\cal L}_s$ contains all terms that
involve soft particles which have energy and momenta of order $mv$.  The terms
we need in the ultrasoft Lagrangian include
\begin{eqnarray} \label{Lu}
  {\mathcal L}_u &=& -{1\over 4}F^{\mu\nu}F_{\mu \nu} + \sum_{\mathbf p}
   \psip p ^\dagger   \Biggl\{ i D^0 - {\left({\bf p}-i{\bf D}\right)^2
   \over 2 m} +\frac{{\mathbf p}^4}{8m^3} \Biggr\} \psip p \nn\\
   & & + \sum_{\mathbf p} \chip p ^\dagger
   \Biggl\{ i D^0 - {\left({\bf p}-i{\bf D}\right)^2
   \over 2 m} +\frac{{\mathbf p}^4}{8m^3} \Biggr\} \chip p \,.
\end{eqnarray}
The covariant derivative is $D^\mu = \partial^\mu + i g \mu_U^\epsilon
A^\mu=(D^0,-\mathbf{D})$, so that $D^0=\partial^0+ig \mu_U^\epsilon A^0$,
${\mathbf D}={\mathbf \nabla}-ig \mu_U^\epsilon {\mathbf A}$, and involves only
the ultrasoft gluon fields. The ultrasoft scale parameter $\mu_U=m\nu^2$, where
$\nu\sim v$ is the subtraction velocity. This $v$ scaling for $\mu_U$ is
required for a consistent power counting in $d$ dimensions \cite{amis2}.  The
covariant derivative on $\psip p$ and $\chip p$ contain the color matrices
$T^A$ and $\bar T^A$ for the $\bf 3$ and $\bf {\bar 3}$ representations,
respectively.

\begin{figure}
  \epsfxsize=10cm \hfil\epsfbox{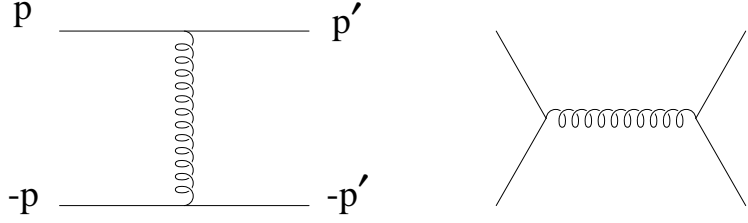}\hfill
{\tighten \caption{QCD diagrams for tree level matching.} \label{fig_tree} }
\end{figure}

The Lagrangian ${\cal L}_p$ includes both the traditional quark potential
and ultrasoft corrections to this potential which we will denote by ${\cal
L}_{pu}$:
\begin{eqnarray}\label{1}
{\mathcal L}_p= - \sum_{\mathbf p,p'} V_{\alpha\beta\lambda\tau}
  \left({\bf p},{\bf p^\prime}\right)\  \mu_S^{2\epsilon}\
  {\psip {p^\prime}}_\alpha^\dagger\:
  {\psip p}_\beta\: {\chip {-p^\prime}}_\lambda^\dagger\:  {\chip {-p}}{}_\tau
  \quad +\ {\cal L}_{pu}\,.
\end{eqnarray}
The terms we need in ${\cal L}_{pu}$ are fixed by reparameterization
invariance~\cite{repar} and will be described below. The on-shell potential
$V({\bf p},{\bf p'})_{\alpha\beta\lambda\tau}$ has an expansion in $\alpha_s$
and $v$, and $\alpha,\beta,\lambda,\tau$ denote color and spin indices. We will
use the color basis in which the potential $V$ is written as a linear
combination of $1 \otimes 1$ and $T \otimes \bar T$.  The tree level diagrams
in Fig.~\ref{fig_tree} generate terms of ${\cal O}(v^{2k} \alpha_s)$ in the QCD
potential.  The order $v^{-2}$ Coulomb potential is
\begin{eqnarray}
 V^{(-2)} &=&  (T^A \otimes \bar T^A) { {\cal V}_c^{(T)} \over {\mathbf k}^2}
 +  (1 \otimes  1) { {\cal V}_c^{(1)} \over {\mathbf k}^2} \,,
\end{eqnarray}
where the coefficients ${\cal V}_c^{(T,1)}$ have an expansion in $\alpha_s$.
The order $v^0$ potential includes
\begin{eqnarray}  \label{V0}
 V^{(0)} &=&  (T^A \otimes \bar T^A) \left[ { {\cal V}_2^{(T)} \over m^2 }
 + { {\cal V}_r^{(T)}\: ({\mathbf p^2 + p^{\prime 2}}) \over 2\, m^2\,
 {\mathbf k}^2}
 + { {\cal V}_s^{(T)} \over m^2}\, {\mathbf S}^2 + {{\cal V}_\Lambda^{(T)}
 \over m^2}\, \Lambda({\mathbf p^\prime ,p}) + { {\cal V}_t^{(T)} \over  m^2}\,
 T({\mathbf k})\right] \nn \\
 && + (1\otimes 1) \left[ { {\cal V}_2^{(1)} \over m^2 } +
 { {\cal V}_s^{(1)} \over m^2}\: {\mathbf S}^2
 \right] \,,
\end{eqnarray}
where ${\bf k} = {\bf p'} - {\bf p}$ and
\begin{eqnarray}
\mathbf S &=& { {\mathbf \bsigma_1 + \bsigma_2} \over 2},
 \qquad \Lambda({\mathbf p^\prime, p }) = -i {\mathbf S \cdot ( p^\prime
 \times p) \over  {\mathbf k}^2 },\qquad
 T({\mathbf k}) = {\mathbf \bsigma_1 \cdot \bsigma_2} - {3\, {\mathbf k
 \cdot \bsigma_1}\,  {\mathbf k \cdot \bsigma_2} \over {\mathbf k}^2} \,.
\end{eqnarray}
Matching the two diagrams in Fig.~\ref{fig_tree} to the $v^{-2}$ and $v^0$
potentials at $\mu=m$ gives
\begin{eqnarray}  \label{bc}
 {\cal V}_c^{(T)} &=& 4 \pi \alpha_s(m)\,, \qquad
 {\cal V}_r^{(T)} = 4 \pi \alpha_s(m)\,, \qquad
 {\cal V}_s^{(T)} = -\frac{4 \pi \alpha_s(m)}{3}+
     {1 \over N_c}\: \pi\, \alpha_s(m) \,, \nn  \\
 {\cal V}_\Lambda^{(T)} &=& -6 \pi \alpha_s(m) \,,\qquad
 {\cal V}_t^{(T)} = -\frac{\pi \alpha_s(m)}{3} \,, \qquad
 {\cal V}_s^{(1)} =  {(N_c^2-1)\over 2N_c^2}\: \pi\, \alpha_s(m) \,,\nn \\
 {\cal V}_c^{(1)} &=& 0 \,,\qquad {\cal V}_2^{(T)} = 0 \,, \qquad
 {\cal V}_2^{(1)} = 0 \,.
\end{eqnarray}
The LL values for the coefficients of the $V^{(-2)}$ and $V^{(0)}$ potentials
will be needed below and are summarized in Appendix~\ref{App_LL}. They are
obtained by using the tree-level matching values in Eq.~(\ref{bc}) and running
using the one loop anomalous dimensions computed in Ref.~\cite{amis}.

The order $1/v$ potential includes
\begin{eqnarray} \label{Lk}
 V^{(-1)} &=& { \pi^2 \over m\,|{\mathbf k}| }\, \Big[
  {\cal V}_k^{(T)} (T^A \otimes \bar T^A) + {\cal V}_k^{(1)} (1\otimes 1)\:
   \Big] \,.
\end{eqnarray}
Tree level matching gives $V^{(-1)}=0$, and $V^{(-1)}$ has zero one loop
anomalous dimension\cite{amis}, so $V^{(-1)}=0$ at LL order as well. At one loop
in QCD, order $1/v$ potentials of the form in Eq.~(\ref{Lk}) are
generated\OMIT{\footnote{\tighten In $d=4-2\epsilon$ the $1/|{\bf k}|$ potential
operator is generated with a $\mu^{2\epsilon}/|{\bf k}|^{1+2\epsilon}$, but we
will define the operator as the value in $d=4$ since we wish to reproduce all
logarithms of $|{\bf k}|$ by the running of the coefficients ${\cal
V}_k^{(T,1)}(\nu)$.}}. The one loop matching of the on-shell potential at $\mu=m$
gives \cite{amis2}
\begin{eqnarray} \label{Lkmatch}
  {\cal V}_k^{(T)} &=& \alpha_s^2(m) \Big( \frac{7 C_A}{8}-\frac{C_d}{8}
  \Big) \,,\qquad
   {\cal V}_k^{(1)} = \alpha_s^2(m) \frac{C_1}{2} \,.
\end{eqnarray}
The color group theory factors are $C_F=(N_c^2-1)/(2N_c)$, $C_A=N_c$,
$C_1=(N_c^2-1)/(4N_c^2)$, and $C_d=N_c-4/N_c$. For the color singlet channel
Eq.~(\ref{Lkmatch}) agrees with Refs.~\cite{Gupta,Yndurain}. In the language of
the threshold expansion\cite{Beneke} the value of the coefficients in
Eqs.~(\ref{bc}) and (\ref{Lkmatch}) are from integrating out off-shell potential
gluons at the hard scale $\mu\sim m$ where $\nu=1$ \cite{amis2}. In the next
section we will compute the two loop anomalous dimension for $V^{(-1)}$, and
determine the NLL value of the coefficients in Eq.~(\ref{Lk}).

In addition we need ultrasoft corrections to the potential which are contained
in ${\cal L}_{pu}$.  Reparameterization invariance\cite{repar} restricts the
form of some of these terms by requiring that only the linear combination
${\bf p} - i{\bf D}$ can appear. Here the covariant derivative acts on a
quark or antiquark field with label ${\bf p}$.  The reparameterization
invariant form of the $T\otimes \bar T$ Coulomb potential operator is
\begin{eqnarray} \label{Lpmult}
  {\Big[ \psip{p^\prime}^\dagger\: T^A {\psip p}\Big]
    \Big[\chip{-p^\prime}^\dagger\: \bar T^A {\chip {-p}}\Big] \over \ppp }
 &\rightarrow& \frac12 \bigg[ \psip{p^\prime}^\dagger\: {T^A \over
  ({\bf p'-p+i\tensor{D}})^2 } \: {\psip p} \bigg]\bigg[
  \chip{-p^\prime}^\dagger\: \bar T^A {\chip {-p}}\bigg] \nn\\
 && + \frac 12 \bigg[ \psip{p^\prime}^\dagger\: T^A \: {\psip p} \bigg]
  \bigg[ \chip{-p^\prime}^\dagger\: {\bar T^A \over
  ({\bf -p'+p+i\tensor{D}})^2}\: {\chip {-p}}\bigg] \,,
\end{eqnarray}
where ${\bf \tensor{D} = \overrightarrow{D} + \overleftarrow{D}}$. Terms in the
$v$ expansion are then generated by expanding Eq.~(\ref{Lpmult}) with ${\bf D}
\ll {\bf p'-p}$. As written the ordering of color generators in
Eq.~(\ref{Lpmult}) is ambiguous. The correct ordering in the expansion is to
write factors of $\overrightarrow{D}$ to the right of the $T^A$, and factors of
$\overleftarrow{D}$ to the left of the $T^A$.  Expanding Eq.~(\ref{Lpmult}) and
keeping only the terms that we will need gives
\begin{eqnarray} \label{Lpu}
  {\cal L}_{pu} &=& {2 i\: {\cal V}_c^{(T)}\, f^{ABC}\over {\mathbf
  k}^4}\mu_S^{2\epsilon}\mu_U^\epsilon \:
    {\mathbf k}\cdot (g {\bf A}^C) \: \psip{p^\prime}^\dagger\:
  T^A {\psip p}\: \chip{-p^\prime}^\dagger\: \bar T^B {\chip {-p}}{}  \nn\\
  && + {\cal V}_c^{(T)}\mu_S^{2\epsilon}\: \psip{p^\prime}^\dagger\:
    \bigg[ \frac{i{\mathbf k}\cdot
    \tensor{\nabla}}{{\mathbf k}^4} -\frac{\tensor{\nabla}^2}{2{\mathbf k}^4}
    +2 \frac{({\mathbf k}\cdot \tensor{\nabla})^2}{{\mathbf k}^6} \bigg]
  T^A {\psip p}\:\chip{-p^\prime}^\dagger\: \bar T^A {\chip {-p}}{} \nn\\
  && + {\cal V}_c^{(T)}\mu_S^{2\epsilon}\: \psip{p^\prime}^\dagger\:
  T^A {\psip p}\:
  \chip{-p^\prime}^\dagger\: \bigg[ \frac{-i{\mathbf k}\cdot
    \tensor{\nabla}}{{\mathbf k}^4} -\frac{\tensor{\nabla}^2}{2{\mathbf k}^4}
    +2 \frac{({\mathbf k}\cdot \tensor{\nabla})^2}{{\mathbf k}^6} \bigg]
    \bar T^A {\chip {-p}}{}  \,,
\end{eqnarray}
where ${\bf \tensor{\nabla} = \overrightarrow{\nabla} +
\overleftarrow{\nabla}}$. The first term couples an ultrasoft gluon to a four
quark operator.
\begin{figure}
  \epsfxsize=2.5cm \hfil\epsfbox{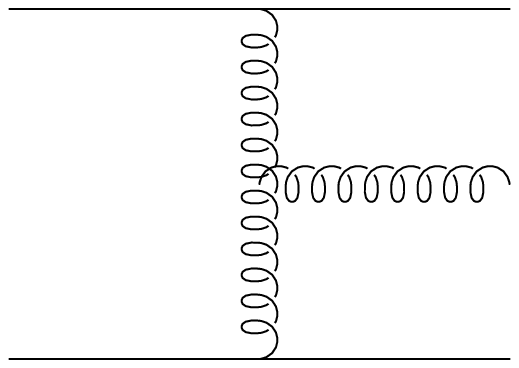}
  \hspace{2.5cm} \lower-25pt \hbox{\Huge $\rightarrow$}
  \epsfxsize=2.5cm \hfil\epsfbox{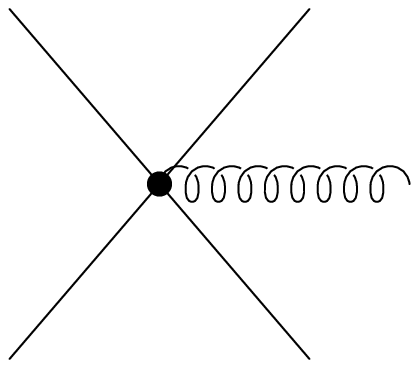}
{\tighten \caption{Matching for the operator attaching an ultrasoft gluon to
a potential interaction.} \label{AV_match} }
\end{figure}
The terms in the second and third lines in Eq.~(\ref{Lpu}) contain order $1/v$
and $v^0$ terms from the multipole expansion of the Coulomb potential.
The first term could also have been determined from the on-shell matching
calculation shown in Fig.~\ref{AV_match}. However, by determining the terms in
Eq.~(\ref{Lpu}) using reparameterization invariance rather than matching we
know that the coefficients remain equal to ${\cal V}_c$ to all orders in
perturbation theory.  This saves us from the extra work that would be involved
in computing the anomalous dimensions for these terms in ${\cal L}_{pu}$.

The terms in the soft Lagrangian include
\begin{eqnarray}\label{Lsoft}
 {\mathcal L}_s &=& \sum_{q} \bigg\{ \abs{q^\mu A^\nu_q -
 q^\nu A^\mu_q}^2 + \bar \varphi_q\, \xslash{q}\, \varphi_q  +
 \bar c_q \, q^2 c_q \ \bigg\}  \\
 && - g^2 \mu_S^{2\epsilon} \sum_{{\mathbf p,p'},q,q'} \bigg\{ \frac12\,
 {\psip {p^\prime}}^\dagger\:
 [A^\mu_{q'},A^\nu_{q}] U_{\mu\nu}^{(\sigma)}\: {\psip p}\: + \frac12\,
 {\psip {p^\prime}}^\dagger\: \{A^\mu_{q'},A^\nu_{q}\} W_{\mu\nu}^{(\sigma)}\:
 {\psip p}\: \nn \\
 && + {\psip {p^\prime}}^\dagger\: [\bar c_{q'},c_{q}] Y^{(\sigma)}\:
 {\psip p}\: + ({\psip {p^\prime}}^\dagger\: T^B Z_\mu^{(\sigma)}\:
 {\psip p} ) \:(\bar \varphi_{q'} \gamma^\mu T^B \varphi_q)  \bigg\}
 + (\psi \to \chi,\: T\to \bar T) \,. \nn
\end{eqnarray}
The fields $A_q^\mu$ and $c_{q}$ are the soft gluon and ghost fields, and
$\varphi_q$ is a massless soft quark field with $n_f$ flavor components. $U$,
$W$, $Y$, and $Z$ are functions of $({\bf p},{\bf p^\prime},q,q')$ and matrices
in spin and the index $\sigma$ denotes the relative order in the $v$ expansion.
For Feynman gauge and the case ${\bf p}^2 = {\bf p'\,}^2$ these functions were
derived in Ref.~\cite{LMR,amis}. Beyond one loop terms proportional to $({\bf
p}^2 - {\bf p'}^2)$ will be needed in ${\cal L}_s$, since besides its soft
energy $q^0\sim m v$ the $A_q$ gluons can carry away a residual energy of order
$m v^2$. The LL values for the functions $U$, $W$, $Y$, and $Z$ can be found in
Appendix~A of Ref.~\cite{amis2}. In addition, some NLL contributions to ${\cal
L}_s$ will be needed and will be discussed in section~\ref{soft_sec}. These
additional contributions are obtained from one loop matching with two-loop
renormalization group improvement.

As an aside, note that it is not necessary to consider the ultrasoft
renormalization of the vertices given in the soft Lagrangian in
Eq.~(\ref{Lsoft}).  One might think that diagrams such as
\begin{eqnarray} \label{ussoft}
\begin{picture}(160,40)(160,1)
   \centerline{
   \epsfxsize=8cm \lower9pt \hbox{\epsfbox{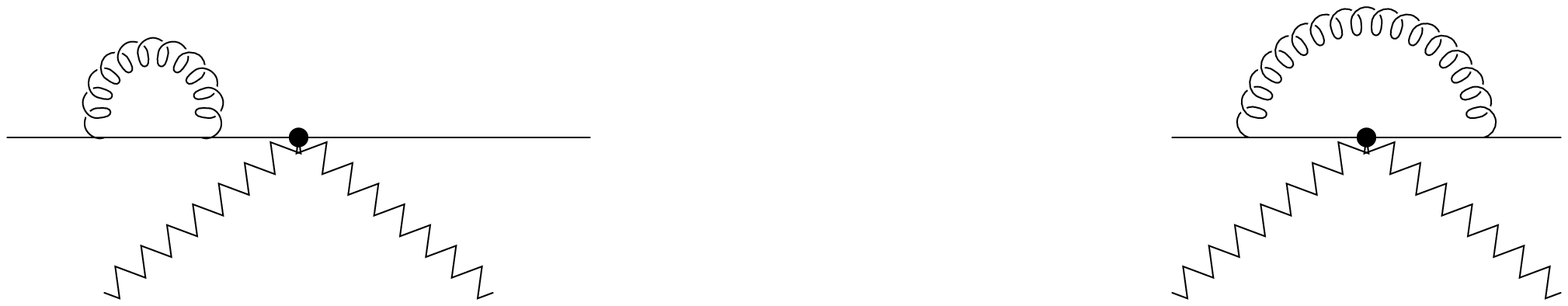}}
   }
\end{picture}
\end{eqnarray}
would effect the running of the coefficients in the soft Lagrangian. Diagrams
analogous to the one in Eq.~(\ref{ussoft}), but with only soft gluons and
quarks generate the running of the coefficient functions in the soft
Lagrangian\footnote{\tighten In Ref.~\cite{amis} the running of ${\cal L}_s$
was calculated by examining loops with soft gluons that contribute to the
Compton scattering process prior to integrating out the soft quarks. There it
was noted that all ultraviolet divergent diagrams are in one-to-one
correspondence with graphs in HQET, so that the running of this Lagrangian
could be obtained from the known running in HQET \cite{run}.} in
Ref.~\cite{amis}. However, the graph in Eq.~(\ref{ussoft}) has only one heavy
quark, so a distinction between soft and ultrasoft gluons is unnecessary at this
point. Noting that the soft vertices will always occur in pairs, it is in fact
consistent to only dress pairs of the soft vertices by ultrasoft gluons:
\begin{eqnarray} \label{ussoft2}
\begin{picture}(160,70)(160,1)
   \centerline{
   \epsfxsize=6cm \lower9pt \hbox{\epsfbox{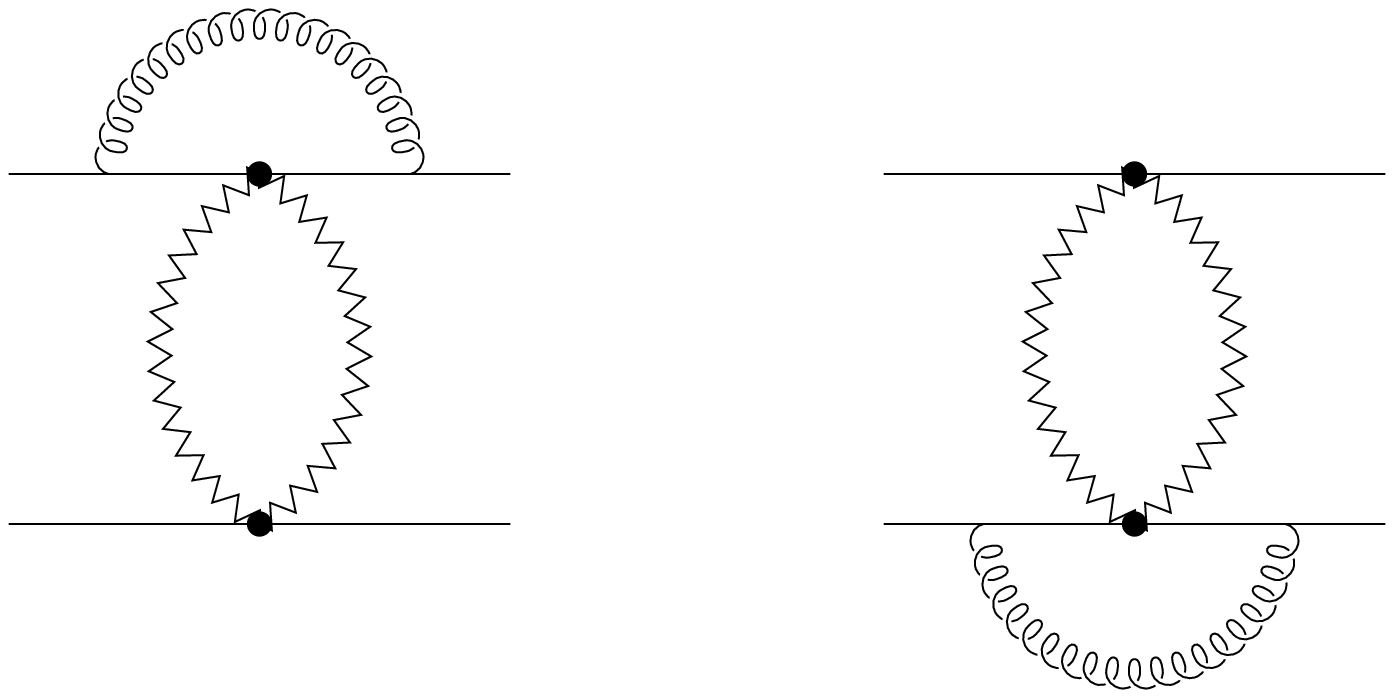}} \qquad\quad
   \epsfxsize=6.3cm \lower0pt \hbox{\epsfbox{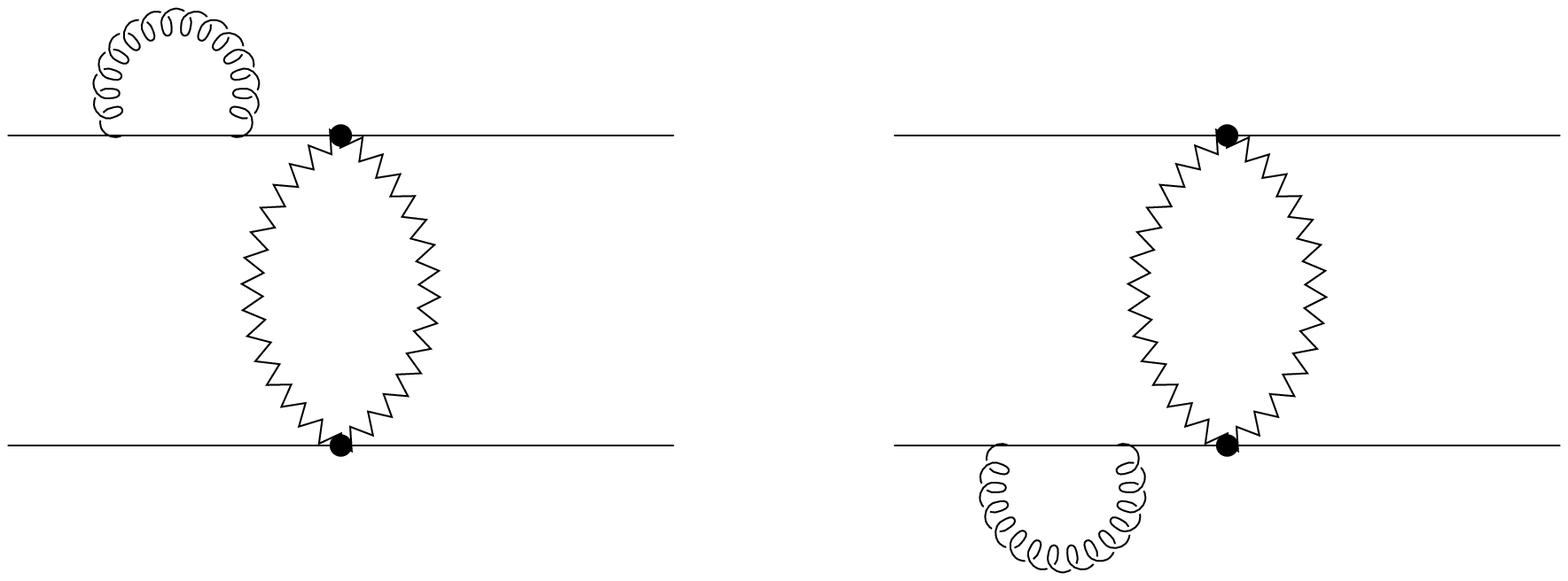}}
   }
\end{picture}\\[-20pt] \nn
\end{eqnarray}
Since these diagrams involve two heavy quarks, both types of gluons can occur.
Since in the end it is only graphs such as Eq.~(\ref{ussoft2}) with two soft
vertices that are relevant for constructing the theory, it is consistent to
include the ultrasoft renormalization of soft vertices as a contribution to the
four-quark operator in Eq.~(\ref{ussoft2}), rather than treating the subgraph
as a contribution to the soft vertex, as in Eq.~(\ref{ussoft}).  Along with the
diagrams in Eq.~(\ref{ussoft2}) there are graphs in which the ultrasoft gluon
is exchanged between the two heavy quarks. 

\newpage
\section{Two loop anomalous dimension for $V^{(-1)}$}

To calculate the NLL anomalous dimension for the $1/|{\bf k}|$ potentials we
need to consider graphs in the effective theory of order\footnote{\tighten This
is the size (in $v$) of the amputated diagrams, so in contrast to the general
power counting formula in Ref.~\cite{LMR} we are not including the powers of $v$
generated by external lines.} $\alpha_s^3/v$. These diagrams come in two
classes, those with soft gluons, and those with a single ultrasoft gluon. The
total anomalous dimension for the $1/|{\bf k}|$ potential is $\gamma^{(1,T)} =
\nu d/d\nu\: {\cal V}_k^{(1,T)}$. Since $\mu_S=m\nu$ and $\mu_U=m\nu^2$,
$\gamma^{(1,T)}$ can be written as the sum of a soft and ultrasoft anomalous
dimension
\begin{eqnarray}
   \gamma^{(1,T)} = \gamma_S^{(1,T)}\,+\, 2\,\gamma_U^{(1,T)} \,,
\end{eqnarray}
where $\gamma_S^{(1,T)} = \mu_S\partial/\partial\mu_S\, {\cal V}_k^{(1,T)}$ and
$\gamma_U^{(1,T)} = \mu_U\partial/\partial\mu_U\, {\cal V}_k^{(1,T)}$.  In the
remainder of this section we discuss the computation of these two loop anomalous
dimensions in detail.  The calculation is split into two parts, graphs with soft
vertices and graphs with ultrasoft vertices.  Graphs with soft gluons only
contribute to $\gamma_S$, while those with an ultrasoft gluon contribute to both
$\gamma_S$ and $\gamma_U$.

\subsection{Soft contributions} \label{soft_sec}

\begin{figure}[t!]
  \epsfxsize=10cm \centerline{\epsfbox{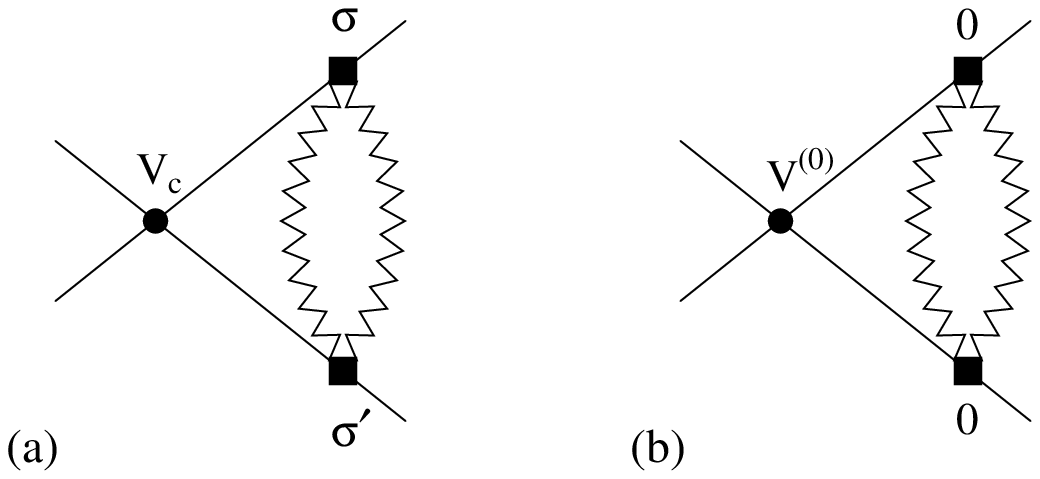}}
  \medskip
  \epsfxsize=14cm \centerline{\epsfbox{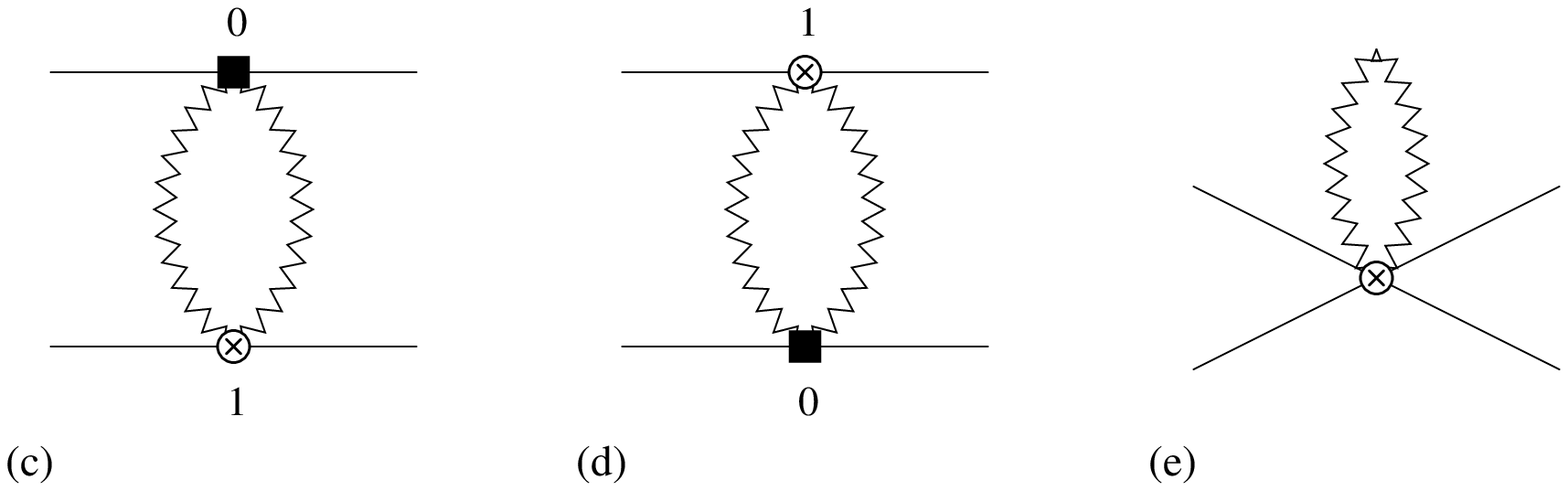}}
  \medskip
{\tighten \caption{Order $\alpha_s^3/v$ graphs containing soft gluons, ghosts,
or massless quarks which are all denoted by a zigzag line. In (a) the dot
denotes the Coulomb potential, while in (b) the dot denotes the order $v^0$
potential.  The boxes denote soft vertices with insertions of the functions
$U^{(\sigma)}$, $W^{(\sigma)}$, $Y^{(\sigma)}$, or $Z^{(\sigma)}$.  In (a) the
indices $\sigma+\sigma'=2$. In (c), (d), and (e) the $\otimes$ vertex is
obtained from the one loop matching in Fig.~\ref{fig_sm}.  } \label{fig_s} }
\end{figure}
The order $\alpha_s^3/v$ diagrams containing soft gluons that contribute to the
anomalous dimension are shown in Fig.~\ref{fig_s}. The sum of diagrams forms
a gauge invariant set.

The two loop graphs in Figs.~\ref{fig_s}a and \ref{fig_s}b involve an iteration
of a potential and a soft loop.  The vertices in these graphs are of LL order
(tree level matching with one loop running) and are given in Ref.~\cite{amis2}.
In Figs.~\ref{fig_s}a and \ref{fig_s}b, counting powers of $v$ from the
propagators and from the loop measures gives a $v^1$, so the sum of powers of
$v$ for the three (amputated) vertices must give an overall $1/v^2$. The $v$
scaling for the two soft vertices is $\sigma+\sigma'-2$. In Fig.~\ref{fig_s} we
show two possibilities: a) has one $V^{(-2)}$ insertion and two soft vertices
from Eq.~(\ref{Lsoft}) such that $\sigma'+\sigma=2$, and b) has one insertion of
a $V^{(0)}$ potential from Eq.~(\ref{V0}) and two soft vertices such that
$\sigma'=\sigma=0$. We could also have a $V^{(-1)}$ potential plus two soft
vertices where $\sigma+\sigma'=1$; however this diagram is identically zero.
The graphs where the potential and soft loop are exchanged simply give a factor
of 2.

\begin{figure}[t!]
  \epsfxsize=9.cm \hfil\epsfbox{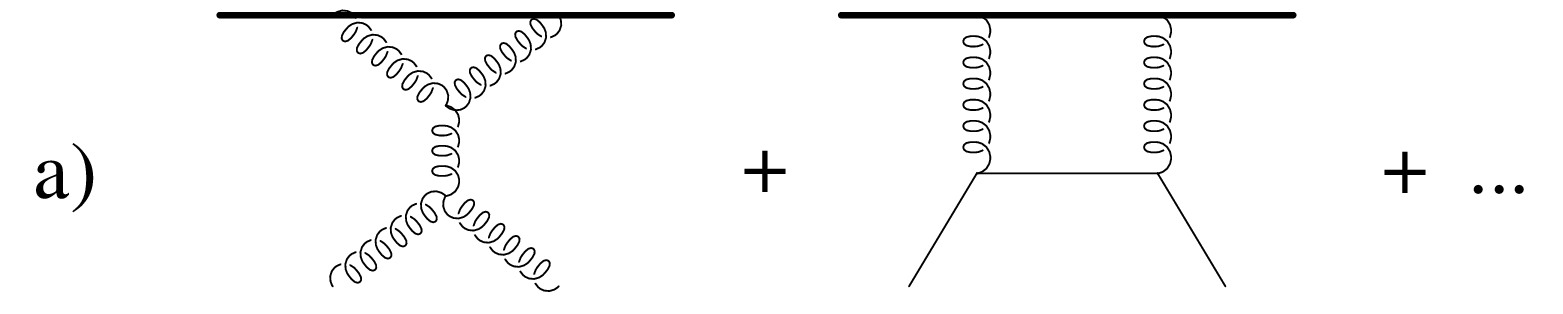}
  \hspace{1.5cm} \lower-20pt \hbox{\Huge $\rightarrow$}
  \epsfxsize=2.5cm \hfil\epsfbox{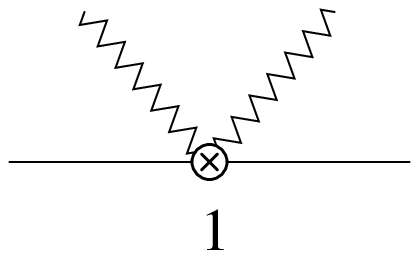} \\[10pt]
  \medskip
  \epsfxsize=9.cm \hfil\epsfbox{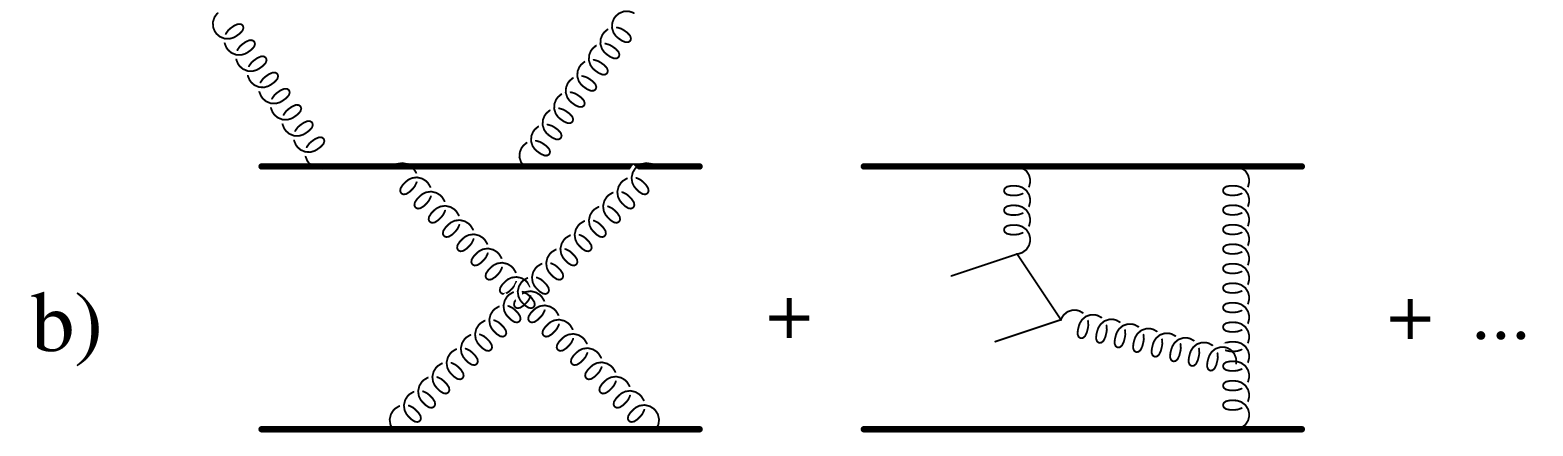}
  \hspace{1.5cm} \lower-25pt \hbox{\Huge $\rightarrow$}
  \epsfxsize=2.5cm \hfil\epsfbox{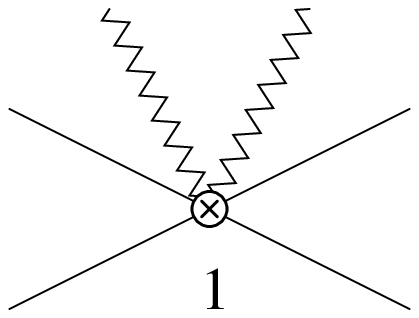}
  \medskip
{\tighten \caption{Contributions to the soft Lagrangian from one loop
matching. In the full theory diagrams on the left the thick lines are massive
quarks, while the thin lines are massless quarks.  The matching in a) gives
$\alpha_s$ corrections to the functions $U^{(1)}$, $W^{(1)}$, $Y^{(1)}$, and
$Z^{(1)}$ in Eq.~(\ref{Lsoft}).  The matching in b) induces new operators that
involve the scattering of soft gluons, ghosts, or light quarks off of a four
fermion potential.  } \label{fig_sm} }
\end{figure}

The one loop graphs in Fig.~\ref{fig_s}c, \ref{fig_s}d, and \ref{fig_s}e involve
additional vertices in ${\cal L}_s$, denoted $\otimes$, which are of NLL order
(from one loop matching with two loop running).  The one loop matching
calculation for these vertices is sketched in Fig.~\ref{fig_sm}.
There are a large number of diagrams in the full theory (graphs on the left hand
side), so only a few representative examples have been shown.  To obtain the
values for the operators on the right hand side we subtract purely soft
effective theory diagrams from those in the full theory. To see how these
operators arise, it is useful to recall that in the threshold
expansion\cite{ls,Gries} soft heavy quarks have a propagator
\begin{eqnarray} \label{sq}
  \frac{1}{q_0+i\epsilon} = {\rm P}\, \frac{1}{q_0} - i\pi \delta(q_0) \,,
\end{eqnarray} 
where ${\rm P}$ stands for the principal value.  In our approach, off-shell
potential gluons and soft quarks are integrated out at the scale $m$ when
constructing the effective theory.  When integrating out the soft heavy quarks
the principal value term in Eq.~(\ref{sq}) goes directly into a coefficient in
the soft Lagrangian since this term is consistent with the scaling in the soft
regime, $q_0\sim mv$.  For instance, in Eq.~(\ref{Lsoft}),
$U_{00}^{(0)}=1/q_0$. The delta function contribution in Eq.~(\ref{sq}) is
associated with the potential regime since $q_0\sim 0$.  When the delta function
appears in a loop in the full theory (or threshold expansion) it forces gluons
in the loop to have zero energy, or in other words to become potential
gluons. It is these contributions which do not appear in the soft effective
theory diagrams and must be made up by the operators shown on the right hand
side of Fig.~\ref{fig_sm}.

The total contribution to the anomalous dimensions from the soft diagrams in
Fig.~\ref{fig_s} is
\begin{eqnarray} \label{sdimk}
 \gamma_{S}^{(T)}&=& -\frac{\beta_0 (7 C_A-C_d)}{8\pi}\: \alpha_s^3(m\nu) 
    - \frac{8 C_A (C_A+C_d)}{3\pi}\: \alpha_s^3(m\nu) \,, \nn \\[10pt]
 \gamma_{S}^{(1)} &=& -\frac{\beta_0 C_1}{2\pi}\: \alpha_s^3(m\nu) 
   + \frac{16 C_A C_1}{\pi}\: \alpha_s^3(m\nu)    \,,
\end{eqnarray}
where $\beta_0=11 C_A/3 - 4 T_F n_f/3$, $n_f$ is the number of massless soft
quarks, and ${\cal V}_c(\nu)= 4\pi\alpha_s(m\nu)$ was used.  In
Eq.~(\ref{sdimk}) the terms proportional to $\beta_0$ can be inferred from
Eq.~(\ref{Lkmatch}).  They simply turn the $\alpha_s(m)$'s in the matching
result into running $\alpha_s$'s. At one and two loops terms proportional to the
$\beta$-function for $\alpha_s$ completely determine the soft anomalous
dimension for the Coulomb potential.  However, the $1/|{\bf k}|$ potentials have
additional contributions because the soft diagrams in Fig.~\ref{fig_s}c,
\ref{fig_s}d, and \ref{fig_s}e have infrared divergences.  The IR divergences
from purely soft diagrams are not true IR divergences in the effective theory.
For instance, in general they do not match up with IR divergences in QCD. The
true IR divergences are from momenta $< mv^2$, and the desired ultraviolet
divergences are from momenta $\ge m$.  Instead, the soft IR divergences are from
momenta $< mv$, and match up with ultrasoft UV divergences that are from momenta
$\ge mv$ to carry these UV divergences up to the hard scale. Writing the soft IR
divergence
\begin{eqnarray}
 \frac{1}{\epsilon_{IR}} = \frac{1}{\epsilon_{UV}} -
  \bigg(\frac{1}{\epsilon_{UV}}-\frac{1}{\epsilon_{IR}} \bigg) \,,
\end{eqnarray}
the first term contributes to the anomalous dimension in Eq.~(\ref{sdimk}). The
$1/\epsilon_{\rm UV} - 1/\epsilon_{\rm IR}$ term can be ignored since it simply
takes the corresponding ultrasoft UV divergence up to the hard scale
$m$.\footnote{\tighten This result, that there are no true IR divergences in
the soft regime has not been proven to all orders in perturbation theory.
However, its is likely that all IR divergences can be attributed to ultrasoft
and collinear gluons in the spirit of the Coleman-Norton theorem~\cite{CN}, plus
IR divergences associated with the Coulomb regime that are reproduced by
iterations of the potential.}  Thus, to compute $\gamma_S$, all divergences from
soft loops should be treated as UV divergences. The terms not proportional to
$\beta_0$ in Eq.~(\ref{sdimk}) can be inferred from the result of the ultrasoft
calculation in Eq.~(\ref{ausdimk}) of the next section.

Despite the fact that Eq.~(\ref{sdimk}) can be inferred without a direct
calculation it is worthwhile to examine the diagrams in Fig.~\ref{fig_s}.
Consider the graphs in Figs.~\ref{fig_s}a and \ref{fig_s}b in Feynman gauge. The
sub-loop with soft gluons is divergent, while the remaining potential loop is
convergent. There is also a set of one loop diagrams (not shown) where the soft
sub-loop is replaced by the one loop counterterms for $V$ derived in
Ref.~\cite{amis}. We find that these counterterm graphs exactly cancel against a
set of divergences in Fig.~\ref{fig_s}a and \ref{fig_s}b. For Fig.~\ref{fig_s}b
there is an exact cancellation, and so there is no operator mixing between
$V^{(0)}$ and $V^{(-1)}$.  However, there are divergences that appear in
Fig.~\ref{fig_s}a that have no corresponding counterterm graphs. Consider the
soft gluon case (the ghost and soft quark cases are similar). After performing
the $k^0$ integration the loop integral for Fig.~\ref{fig_s}a is
\begin{eqnarray} \label{int1}
  \int\!d^{d-1}{\mathbf q}{1 \over (E-{\mathbf q}^2/m) ({\mathbf p-q})^2 }
  \Bigg[ \int\!d^dt {U_{\mu\nu}^{(\sigma)}\: U^{\mu\nu\,(\sigma')} \over
  t^2 [(t^0)^2 -({\mathbf t+q-p'\,})^2] } \Bigg] \,,
\end{eqnarray}
where ${\bf p}$ and ${\bf p'}$ are the momenta of the incoming and outgoing
quarks, $E={\bf p^2}/m$ and in Eq.~(\ref{int1}) the $U$'s depend on ${\bf q}$,
${\bf t}$, ${\bf p'}$ and $t^0$. For the divergences of interest performing the
$t$ integral (the soft loop) in $d=4-2\epsilon$ dimensions gives a factor of
\begin{eqnarray} \label{offsub}
  { ({\bf p'\,^2-q^2})^2 \over m^2({\bf p'-q})^{4+2\epsilon}\ \epsilon } + 
  \ldots \,,
\end{eqnarray}
where the ellipsis denote order $\epsilon^0$ terms. The remaining loop
integration is finite.  For the one loop graph that corresponds to the soft
sub-loop, the loop momenta ${\bf q}$ in Eq.~(\ref{offsub}) is replaced by ${\bf
p'}$, and this diagram vanishes on-shell by energy conservation so no
counterterm is generated. Performing the final integration and including the
factor of $2$ from the left-right symmetric graph gives
\begin{eqnarray} \label{Fig3a}
  \mbox{Fig.}~\ref{fig_s}a\  =\ i\, {\cal V}_c^{(T)}(\mu_S)\:
  \alpha_s^2(\mu_S)\,
   { \beta_0\, \mu_S^{2\epsilon} \over 32\: m |{\bf k}|  } \:
  (T^A T^B \otimes \bar T^A \bar T^B)
  \bigg[ \frac{1}{\epsilon} + 2\ln\Big(\frac{\mu_S^2}{|{\bf k}|^2}\Big)
  + \ldots \bigg]  \,.
\end{eqnarray}

The divergence in Eq.~(\ref{Fig3a}) contributes to the anomalous dimension for
${\cal V}_k$.  This seems slightly unusual because it was generated by a
sub-divergence, whereas usually only the overall divergence in a diagram is
relevant. In this diagram the loop integral with soft gluons generates a
$1/\epsilon$ pole, and the remaining potential loop integral generates the
$1/|{\bf k}|$ factor. The corresponding diagrams in QCD include graphs such as
the vacuum polarization of one of the gluons in the box diagram.  In this full
theory graph the subdivergence due to the vacuum polarization insertion would be
canceled by a counterterm. However, in the effective theory this divergence is
instead absorbed into the coefficients of terms in the quark potential because
the potential gluon components have been integrated out. Since matching the full
theory box diagram gives a contribution to the $1/|{\bf k}|$ potential, it is
not surprising that gluon vacuum polarization in the box graph contributes to
the anomalous dimension of this potential.  An alternative to the approach used
to derive Eq.~(\ref{Fig3a}) is to use off-shell matching and running.  In this
case the soft anomalous dimension analysis will be different (since, for
instance, the off-shell potential is gauge dependent).  This approach is
discussed in Appendix~\ref{App_off}, where it is shown that the final answer for
physical observables is unchanged.

The diagrams in Figs.~\ref{fig_s}c, \ref{fig_s}d, and \ref{fig_s}e are tedious
to evaluate due to the large number of diagrams necessary for the matching
calculation in Fig.~\ref{fig_sm}.\footnote{\tighten In fact once a contribution
to $\otimes$ has been identified it is simpler to directly evaluate the two-loop
diagram obtained by combining the steps in Fig.~\ref{fig_sm} and
Fig.~\ref{fig_s}.}  Since the graphs in Figs.~\ref{fig_s}c, \ref{fig_s}d, and
\ref{fig_s}e are one-loop diagrams there is no cancellation from counterterms.
In Feynman gauge
\begin{eqnarray} \label{Fig3cde}
 \mbox{Fig.}~\ref{fig_s}c,d,e\ &= & i\, \frac{\pi^2\mu_S^{2\epsilon}}
  {m|{\bf k}|} \frac{\alpha_s^3(\mu_S)} {4\pi\: \epsilon} \bigg\{ \frac43\, 
  C_1 T_F n_f (1 \otimes  1)+\Big( C_A T_F n_f-\frac{1}{3}\, C_d T_F n_f \Big) 
  (T^A \otimes \bar T^A) \nn\\*
  &&\qquad\qquad\qquad\ \ + \frac{37}{3}\, C_A C_1 (1 \otimes  1) - 
  \Big(\frac{7}{4} \,C_A C_d + \frac{65}{12}\,  C_A^2\Big) (T^A \otimes 
  \bar T^A) \bigg\} \,.
\end{eqnarray}
Only the terms proportional to $n_f$ in Eq.~(\ref{Fig3cde}) have been checked by
direct calculation. Also, in Eq.~(\ref{Fig3cde}) we have assumed that the
$\otimes$ operators simply run with the strong coupling constant,
$\alpha_s(m\nu)$. To motivate this, recall that a theory of propagating soft
quarks and gluons has the same singularity structure as Heavy Quark Effective
Theory (HQET).  The operators that are generated in Fig.~\ref{fig_sm} correspond
to diagrams in HQET with insertions of $\nabla^2/(m)$ or a ${\bf p}\cdot {\bf
A}/m$ vertex. Since these operators do not run in HQET the operators that are
constructed in Fig.~\ref{fig_sm} should also simply run with the strong coupling
constant.

\subsection{Ultrasoft contributions}

Next consider the order $\alpha_s^3/v$ diagrams with ultrasoft gluons. At
one loop we have a single insertion of the $V^{(-1)}$ potential dressed by an
ultrasoft gluon with $gA^0$ couplings. These diagrams are obviously zero in
Coulomb gauge and in Feynman gauge it was shown in Ref.~\cite{LMR} that, with
any potential, the sum of this set of one loop diagrams is also identically
zero. At two loops in an arbitrary gauge there are many possible
contributions. There are diagrams with two $V^{(-2)}$ vertices, and an ultrasoft
tadpole generated by the seagull ${\bf A^2}$ operator attached to one of the
quark lines. These graphs do not have logarithmic divergences and therefore do
not contribute to the anomalous dimension.  There are also diagrams that are
zero because they are odd in an ultrasoft momenta which we can omit. Next 
consider the topologies shown in Fig.~\ref{fig_us1}.  Several
classes of diagrams are generated depending on the vertices used:
\begin{figure}
  \centerline{
    \epsfxsize=3.5cm \epsfbox{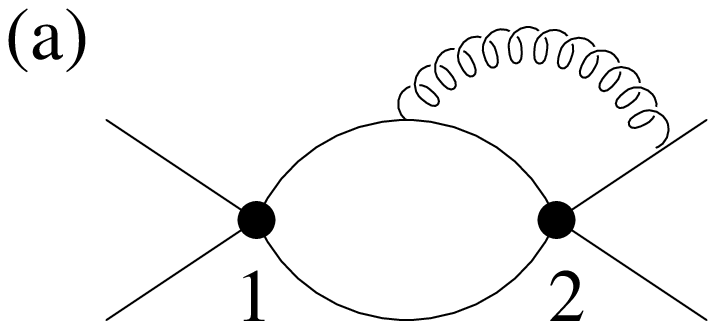} \qquad
    \epsfxsize=3.5cm \epsfbox{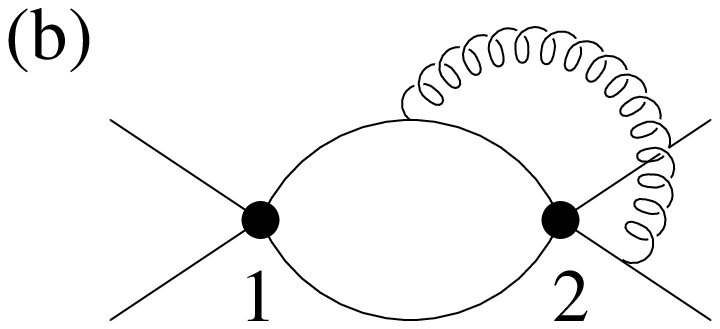} \qquad
    \epsfxsize=3.5cm \epsfbox{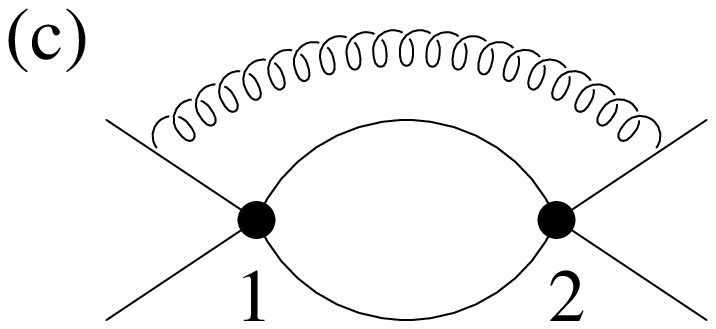} \qquad
    \epsfxsize=3.5cm \epsfbox{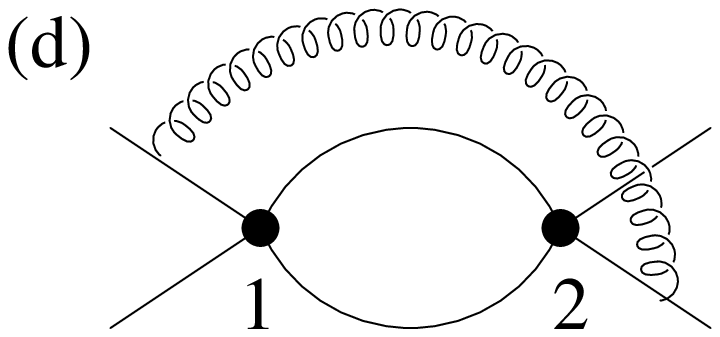} } \vspace{0.3cm}
  \centerline{
    \epsfxsize=3.5cm \epsfbox{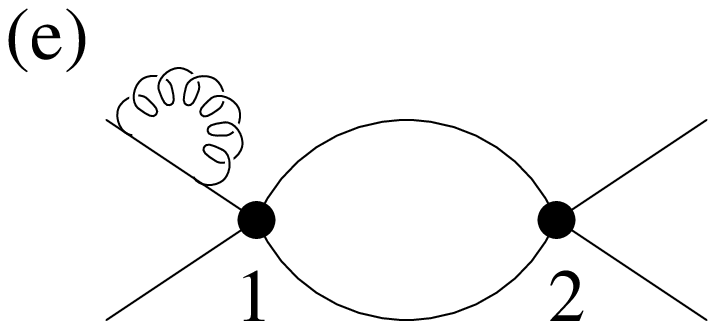} \qquad
    \epsfxsize=3.5cm \epsfbox{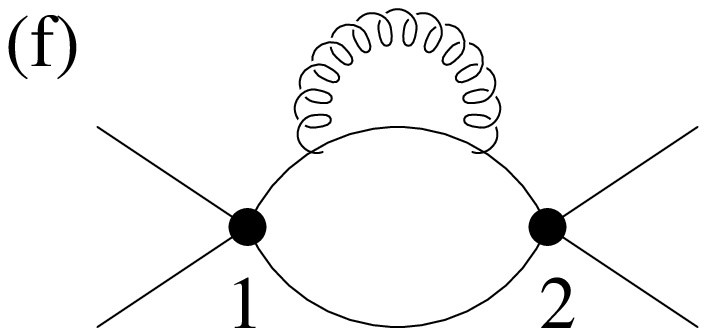} \qquad
    \epsfxsize=3.5cm \epsfbox{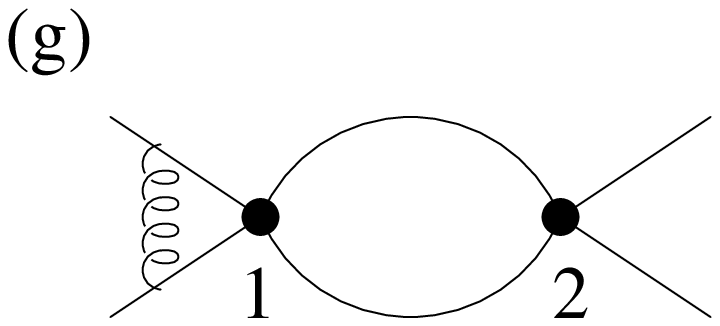} \qquad
    \epsfxsize=3.5cm \epsfbox{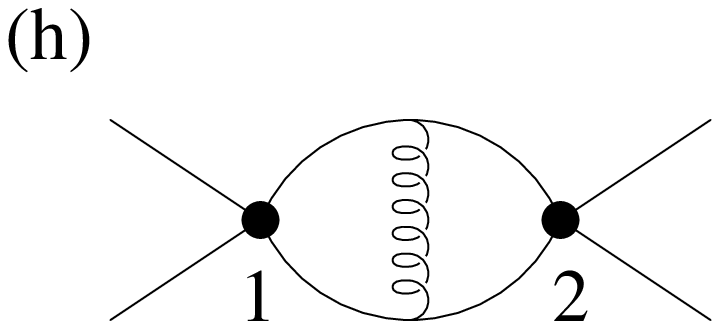} }
{\tighten \caption{Order $\alpha_s^3/v$ two loop graphs with an ultrasoft gluon
and two potential insertions. The topologies shown give several different
classes of diagrams depending on the vertices used as explained in the text.
The diagrams obtained by flipping the graphs left-to-right and up-to-down are
to be understood.} \label{fig_us1} }
\end{figure}

{\tighten \begin{enumerate}
\item vertex $1$ and $2$ are $V^{(-2)}$, and the ultrasoft gluon couples with
  ${\bf p \cdot A}/m$,
\item vertex $1$ and $2$ are $V^{(-2)}$, the ultrasoft gluon couples with
  $gA^0$, and there are two insertions of ${\bf p\cdot\nabla}/m$ on quark lines,
\item vertex $1$ and $2$ are $V^{(-2)}$, the ultrasoft gluon couples with
  $gA^0$, and there is one insertion of ${\bf \nabla^2}/m$ on a quark line,
\item vertex $1$ and $2$ are $V^{(-2)}$, the ultrasoft gluon couples with
  $gA^0$, and there is one insertion of ${\bf p^4}/m^3$ on a quark line (together
  with the expansion of factors of the energy that appear in lower order diagrams
  which should be viewed as corrections to the effective theory states, see
  Refs.~\cite{ls,amis2}),
\item vertex $1$ is $V^{(-2)}$, vertex $2$ is $V^{(0)}$ and the ultrasoft gluon
  couples with $g A^0$,
\item vertex $1$ is the order $1/v$ potential from the multipole expansion of the
 Coulomb potential given in Eq.~(\ref{Lpu}), vertex $2$ is $V^{(-2)}$, and
 the ultrasoft gluon couples with $gA^0$ vertices.
\item vertex $1$ is the order $v^0$ potential in Eq.~(\ref{Lpu}), vertex $2$
 is $V^{(-2)}$, there is one insertion of ${\bf p \cdot \nabla}/m$ on a quark
 line, and the ultrasoft gluon couples with $gA^0$ vertices.
\item vertex $1$ and $2$ are $v^0$ potentials from Eq.~(\ref{Lpu}), and the
 ultrasoft gluon couples with $gA^0$ vertices.
\end{enumerate} }
\noindent Insertions of operators with $\nabla$'s only need to be considered on
quark propagators where the multipole expansion was used; together with the
graphs in cases 6, 7, and 8 they build up the sub-leading terms in the
multipole expansion.

The graphs in cases 1, 2, and 5 do not give any contribution to the two loop
anomalous dimension. The reason is that all ultraviolet divergences are exactly
canceled by one loop counterterm graphs (these counterterms are generated by one
loop graphs having one potential insertion dressed by an ultrasoft gluon, and in
Feynman gauge were calculated in Refs.~\cite{LMR,amis}).  The graphs with the
topology in Fig.~\ref{fig_us1}c,d are ultraviolet finite.  For the remaining
diagrams we can identify the corresponding counterterm graph by simply shrinking
the smallest loop containing the ultrasoft gluon to a point.  The only subtlety
occurs in Fig.~\ref{fig_us1}h which has overlapping potential and ultrasoft loop
integrals.  For heavy scalars this graph was analyzed in Ref.~\cite{Beneke} with
the threshold expansion, while in an effective theory for heavy fermions this
diagram was analyzed in Ref.~\cite{ms2} for the case where the ultrasoft gluon
is massive and couples with derivatives at the vertices.  Since the nature of
the overlapping integrals does not depend on the structure of the numerator or
on having a massive gluon, this analysis will not be repeated here. In the
effective theory this diagram comes with the appropriate factor of $2$, so that
it is canceled by the two one loop counterterm graphs that correspond to either
making the loop integral on the right large {\em or} making the loop integral on
the left large.  Note that for a $\phi^3$ relativistic theory in $d=6$
\cite{Collins}, the sub-divergences for this diagram are also canceled in this
way, but leave an overall divergence.  For the non-relativistic effective theory
diagram this overall divergence is not present.

For the graphs in case 4, the sum of insertions on quark lines inside the loop
shared by the ultrasoft gluon are finite.  The graphs with insertions on quark
lines outside this loop are ultraviolet divergent, but are exactly canceled by
the counterterm diagrams that correspond to shrinking the ultrasoft loop to a
point.  Therefore, the graphs in case 4 also do not contribute to the anomalous
dimension.

In a general gauge the graphs in cases 3, 6, 7, and 8 will contribute to the
anomalous dimension. There are also additional graphs with the vertex in
Eq.~(\ref{Lpmult}) which involves the coupling of an ultrasoft ${\bf A}^i$ gluon
to a potential vertex. Together this set of graphs gives a gauge invariant
contribution to the anomalous dimension, so we can simplify their computation by
choosing the most convenient gauge. We will choose Coulomb gauge since the
graphs in cases 3, 6, 7, and 8 involve ultrasoft $A^0$ gluons and vanish in this
gauge.  The result for the infinite part of the the remaining diagrams in
Coulomb gauge is:\\[-30pt]
\begin{mathletters} \label{Cusoft}
\begin{eqnarray}
\begin{picture}(90,40)(-5,1)
   \epsfxsize=2.6cm \lower9pt \hbox{\epsfbox{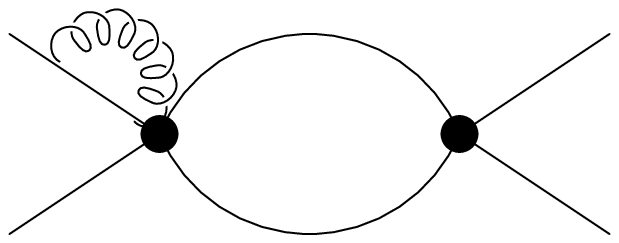}}
\end{picture} &=& 0 \,, \\
\begin{picture}(90,40)(-5,1)
   \epsfxsize=2.6cm \lower9pt \hbox{\epsfbox{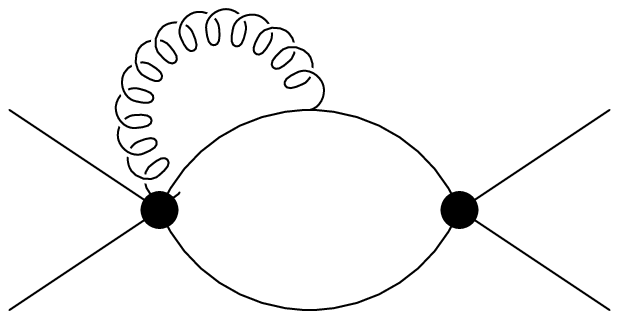}}
\end{picture} &=&
 \frac{ -i\, \alpha_s(\mu_U\!)\mu_S^{2\epsilon}\: [  {\cal V}_c^{(T)} ]^2 }
   {8 \pi\, m\, |{\bf k}|\, \epsilon} \bigg[ C_A C_1\: 1\otimes 1 -\frac14 C_A
   (C_A + C_d) T^A \otimes \bar T^A \bigg] \,, \\
\begin{picture}(90,40)(-5,1)
   \epsfxsize=2.6cm \lower9pt \hbox{\epsfbox{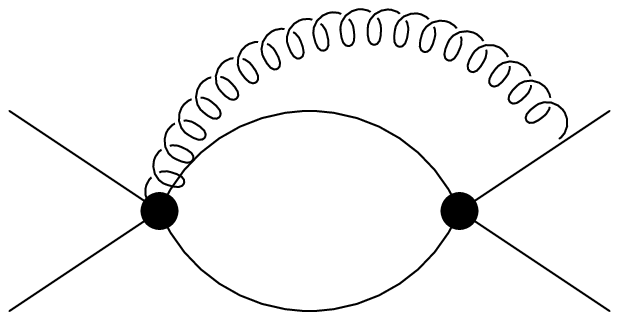}}
\end{picture} &=&
 \frac{ -i\, \alpha_s(\mu_U\!)\mu_S^{2\epsilon}\: [  {\cal V}_c^{(T)} ]^2 }
   {8 \pi\, m\, |{\bf k}|\, \epsilon} \ \frac{C_A C_1}{3}\ 1\otimes 1 \,, \\
\begin{picture}(90,40)(-5,1)
   \epsfxsize=2.6cm \lower9pt \hbox{\epsfbox{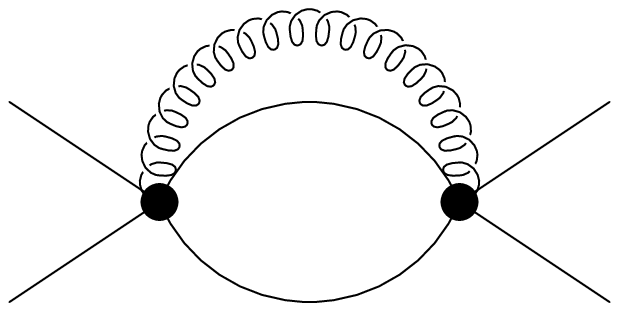}}
\end{picture} &=&
 \frac{ -i\, \alpha_s(\mu_U\!)\mu_S^{2\epsilon}\: [  {\cal V}_c^{(T)} ]^2 }
   {8 \pi\, m\, |{\bf k}|\, \epsilon} \bigg[ \frac{2}{3} C_A C_1\: 1\otimes 1
   -\frac{1}{12} C_A (C_A + C_d) T^A \otimes \bar T^A \bigg] \,. \\[-15pt] \nn
\end{eqnarray}
\end{mathletters}
Equations~(\ref{Cusoft}b,c) include the factor of four from their left-right and
up-down mirror graphs. In the graphs in Eq.~(\ref{Cusoft}) the factor
\begin{equation}
  {\mu_U^{2\epsilon}\mu_S^{4\epsilon} \over \epsilon \abs{\bf k}^{1+2\epsilon}
  E^{2\epsilon}} = {\mu_S^{2\epsilon}\over |{\bf k}|} \left[{1\over\epsilon}+
  \ln\bigg({\mu_S^2\over\abs{\bf k}^2}\bigg) + \ln\bigg({\mu_U^2 \over E^2}\bigg)
  \right]
\end{equation}
occurs, where $E={\bf p^2}/m$. The first term is included in the result
displayed in Eq.~(\ref{Cusoft}), and the remaining terms are not shown. For
$\nu\sim v$, we have $\mu_S \sim \abs{k} \sim m v$, $\mu_U\sim E\sim mv^2$, so
the soft and ultrasoft scale factors correctly minimize possible large
logarithms in the effective theory.

From the sum of diagrams in Eq.~(\ref{Cusoft}) we have the following result for
the soft and ultrasoft anomalous dimensions of the order $1/|{\bf k}|$
potentials:
\begin{eqnarray} \label{ausdimk}
 \gamma_{S}^{(T)} = \gamma_{U}^{(T)} &=&
    C_A (C_A+C_d)\: \frac{{\alpha_s(m\nu^2)} \: [{\cal V}_c^{(T)}(\nu)]^2 }
    {12 \pi^3} \,, \nn \\[10pt]
 \gamma_{S}^{(1)} = \gamma_{U}^{(1)} &=&
   -C_A C_1\:\frac{ {\alpha_s(m\nu^2)}\: [{\cal V}_c^{(T)}(\nu)]^2 }{2 \pi^3}
  \,.
\end{eqnarray}
Recall that the soft and ultrasoft anomalous dimensions are given by
differentiating with respect to $\ln\mu_S$ and $\ln\mu_U$ respectively.

It is interesting to ask how the result in Eq.~(\ref{ausdimk}) would be
reproduced if terms that vanish by the equations of motion were included in our
effective Lagrangian. It is possible to include a term in our
effective Lagrangian which makes the graph in Eq.~(\ref{Cusoft}b) give no
contribution to the anomalous dimension, but the final result for observables
remains invariant. This example is discussed in Appendix~\ref{App_off}.

\section{Results}

In this section the NLL heavy quark $1/|{\bf k}|$ potential and production
current are discussed.  After presenting the renormalization group improved
results, we re-expand to compare to the finite order results in
Refs.~\cite{Brambilla} and \cite{Penin}. We also discuss the behavior of the NLL
$1/|{\bf k}|$ potential and the NLL production current as we run down from
$\nu=1$ to $\nu=v$, where $v$ is the Coulombic velocity.

\subsection{The NLL \lowercase{$1/|{\bf k}|$} potentials} \label{nll1}

The total soft and ultrasoft anomalous dimensions for the on-shell $1/|{\bf k}|$
potentials are obtained by adding Eqs.~(\ref{sdimk}) and (\ref{ausdimk}). Using
the LL coefficient for the Coulomb potential, ${\cal V}_c(\nu)=
4\pi\alpha_s(m\nu)$, we find
\begin{eqnarray} \label{totadsus}
\gamma_S^{(T)} &=& - \frac{1}{8\pi} \beta_0 (7 C_A-C_d) \: {[\alpha_s(m\nu)]^3}
  - \frac{8}{3\pi} C_A (C_A+C_d)\: {[\alpha_s(m\nu)]^3} \nn\\ 
  && + \frac{4}{3\pi} C_A (C_A+C_d)\: {\alpha_s(m\nu^2) [\alpha(m\nu)]^2}
  \,, \nn \\[5pt]
\gamma_S^{(1)} &=& - \frac{1}{2\pi} \beta_0 \,C_1\: {[\alpha_s(m\nu)]^3}
  + \frac{16}{\pi}\, C_A C_1\: {[\alpha_s(m\nu)]^3} 
  -\frac{8}{\pi}\, C_A C_1\: {\alpha_s(m\nu^2) [\alpha_s(m\nu)]^2 }
  \,, \nn\\[5pt]
\gamma_U^{(T)} &=&  \frac{4}{3\pi} C_A (C_A+C_d)\: {\alpha_s(m\nu^2)
  [\alpha_s(m\nu)]^2 } \,, \nn \\[5pt]
\gamma_U^{(1)} &=&  -\frac{8}{\pi}\, C_A C_1\: {\alpha_s(m\nu^2)
  [\alpha_s(m\nu)]^2 } \,.
\end{eqnarray}
In full QCD (taking the QCD scale parameter $\mu=m$), the first logarithm
generated by the ultrasoft anomalous dimension corresponds to a $\ln(E/m)$,
while the first logarithm generated by the soft anomalous dimension corresponds
to a $\ln(|{\bf k}|/m)$.  For the velocity renormalization group, the total
anomalous dimension is simply $\gamma_S +2 \gamma_U$:
\begin{eqnarray} \label{totad}
\nu {\partial \over \partial\nu} {\cal V}_k^{(T)}(\nu) &=&
  - \frac{1}{8\pi} \beta_0 (7 C_A-C_d) \: {[\alpha_s(m\nu)]^3}
  - \frac{8}{3\pi} C_A (C_A+C_d)\: {[\alpha_s(m\nu)]^3} \\ 
  && + \frac{4}{\pi} C_A (C_A+C_d)\: {\alpha_s(m\nu^2) [\alpha(m\nu)]^2}
  \,, \nn \\[5pt]
\nu {\partial \over \partial\nu} {\cal V}_k^{(1)}(\nu) &=&
   - \frac{1}{2\pi} \beta_0 \,C_1\: {[\alpha_s(m\nu)]^3}
  + \frac{16}{\pi}\, C_A C_1\: {[\alpha_s(m\nu)]^3} 
  -\frac{24}{\pi}\, C_A C_1\: {\alpha_s(m\nu^2) [\alpha_s(m\nu)]^2 } \nn \,.
\end{eqnarray}
Note that no other $1/v$ potentials are generated for $\nu<1$ by operator
mixing. Integrating Eq.~(\ref{totad}) using the one loop $\beta$-function
for $\alpha_s$ and the one loop boundary condition in Eq.~(\ref{Lkmatch})
gives
\begin{eqnarray} \label{Vksoln}
  {\cal V}_k^{(T)}(\nu)  &=& \frac{(7 C_A-C_d)}{8}\: \alpha_s^2(m)
   + \left[\frac{(7C_A-C_d)}{8}+\frac{8\,C_A(C_A+C_d)}{3\beta_0}\right]
   ( z^2 - 1 )\: \alpha_s^2(m)  \\
   &&+ \frac{8 C_A(C_A+C_d)}{\beta_0} \Big[ z-1-2\ln(w)\Big]\:\alpha_s^2(m)
   \,, \nn\\
  {\cal V}_k^{(1)}(\nu) &=& \frac{C_1}{2} \: \alpha_s^2(m)
   +\left[ \frac{C_1}{2} -\frac{16\,C_A C_1}{\beta_0}\right] (z^2 - 1)\:
   \alpha_s^2(m) 
   - \frac{48 C_A C_1}{\beta_0}\: \alpha_s^2(m) \Big[ z-1-2\ln(w) \Big] \,,\nn
\end{eqnarray}
where
\begin{eqnarray}
  z=\frac{ \alpha_s(m\nu) }{ \alpha_s(m) }\,,\qquad\quad
  w=\frac{ \alpha_s(m\nu^2) }{ \alpha_s(m\nu) }={1\over 2-z} \,.
\end{eqnarray}
Projecting onto the color singlet channel, ${\cal V}^{(s)}={\cal V}^{(1)} - C_F
{\cal V}^{(T)}$, and setting $C_d=8 C_F-3 C_A$ and $C_1=C_A C_F/2-C_F^2$ gives
\begin{eqnarray} \label{Vkssoln}
  {\cal V}_k^{(s)}(\nu)  &=& \Big(\frac{C_F^2}{2}-C_F C_A \Big)\: \alpha_s^2(m)
  + \left[ \frac{C_F^2}{2} -C_F C_A - \frac{8 C_A C_F (C_A+2 C_F)}{3\beta_0}
  \right](z^2-1)\: \alpha_s^2(m) \nn\\
  && - \frac{8 C_A C_F (C_A+2 C_F)}{\beta_0}\Big[ z-1-2\ln(w) \Big]\: 
  \alpha_s^2(m) \,.
\end{eqnarray}
The projection onto the color octet channel is ${\cal V}_k^{(o)}={\cal
V}_k^{(1)} + (C_A/2-C_F) {\cal V}_k^{(T)}$.

Logarithmic corrections to the color singlet $V^{(-1)}$ potential were also
considered by Brambilla et al.~\cite{Brambilla} using pNRQCD, but were not
resummed. Brambilla et al.\ have
\begin{eqnarray} \label{VkB}
  {\cal V}_k^{(s)}(\nu) &=& -C_F C_A \alpha_s^2(r)-\frac{4 C_A C_F (C_A+2 C_F)}
  {3\pi}\: {\alpha_s(\mu) \alpha_s(r)^2} \ln(\mu r) \,.
\end{eqnarray}
To compare this expression to ours: $\alpha(r)\to \alpha(m\nu)$,
$r\to 1/(m\nu)$ and $\mu=\mu_U=m\nu^2$ since $r$ corresponds to the soft scale
and $\mu$ corresponds to the ultrasoft scale.  Expanding Eq.~(\ref{VkB}) in
$\alpha_s(m)$ gives
\begin{eqnarray} \label{pertB}
 {\cal V}_k^{(s)}(\nu)  &=& -C_F C_A \, \alpha_s^2(m) 
 +  \frac{\beta_0 C_F C_A}{\pi}  \alpha_s^3(m)\ln[1/(mr)] \nn\\ 
 && -\frac{4 C_A C_F (C_A+2 C_F)}{3\pi}\: {\alpha_s^3(m)}\ln(\mu r)
  + \ldots \,. 
\end{eqnarray}
To compare this to our result we expand the resummed logarithms in
Eq.~(\ref{Vkssoln}):
\begin{eqnarray} \label{pert1}
 {\cal V}_k^{(s)}(\nu) &=& \Big( \frac{C_F^2}{2}-C_F C_A \Big) \alpha_s^2(m)  
  +\frac{\beta_0\,C_F C_A}{\pi}\: \alpha_s^3(m)\ln(\nu)
     -\frac{\beta_0\,C_F^2}{2\pi}\: \alpha_s^3(m)\ln(\nu)\nn  \\[5pt]
 &&  -\frac{4 C_A C_F (C_A+2 C_F)}{3\pi}\: {\alpha_s(m)^3}\ln(\nu) +\ldots 
  \,. 
\end{eqnarray}
In Eq.~(\ref{pert1}), the first and second $\ln(\nu)$ terms are entirely from
the soft anomalous dimension in Eq.~(\ref{totadsus}), while the third $\ln(\nu)$
term is from a combination of the ultrasoft and soft anomalous dimensions. The
first $\ln(\nu)$ term in Eq.~(\ref{pert1}) agrees with the
$\ln[1/(mr)]=\ln(\nu)$ term in Eq.~(\ref{pertB}), and the third $\ln(\nu)$ term
in Eq.~(\ref{pert1}) agrees with the $\ln(\mu r)=\ln(\nu)$ term in
Eq.~(\ref{pertB}).

The second term in Eq.~(\ref{pert1}) does not appear in Brambilla et al.'s
expression in Eq.~(\ref{pertB}). This is because it depends on whether an
on-shell or off-shell potential is used for the matching and
running\footnote{\tighten The issue of on-shell versus off-shell potential was
discussed in Ref.~\cite{amis2}, and in the context of the leading log summation
is discussed further in Appendix~\ref{App_off}.}. We have used an on-shell
potential, while Ref.~\cite{Brambilla} uses off-shell Coulomb gauge which
includes a ${\cal V}_{\Delta 2}^{(T)}=-4\pi\alpha_s(r)$ potential as defined in
Eq.~(\ref{V0a}) of Appendix~B. The second logarithm in Eq.~(\ref{pert1}) should
appear from the coefficient of this potential.  Transforming Brambilla et al.'s
${\cal V}_{\Delta 2}^{(T)}$ potential to a ${\cal V}_k^{(s)}$ potential using
Eq.~(\ref{trnsfm2}) gives
\begin{eqnarray} \label{pertB2}
 {\cal V}_k^{(s)}(\nu) &=& \frac{C_F^2}{2}\, \alpha_s(r)^2 \nn\\[5pt]
  &=& \frac{C_F^2}{2}\alpha_s^2(m) - \frac{\beta_0 C_F^2}{2\pi}\, 
  \alpha_s^3(m)\, \ln[1/(mr)] +\ldots \,.
\end{eqnarray}
The term with a $\ln[1/(mr)]=\ln(\nu)$ agrees with the second $\ln(\nu)$ term in
Eq.~(\ref{pert1}).  The sum of terms without logarithms in Eqs.~(\ref{pertB})
and (\ref{pertB2}) also agrees with Eq.~(\ref{pert1}). Note that the next term
in the series in Eq.~(\ref{pert1}) does not give agreement with
Eqs.~(\ref{pertB}) and (\ref{pertB2}) since Brambilla et al. did not attempt to
systematically sum all the logarithms.

\begin{figure}[!t]
  \centerline{\epsfxsize=10.0truecm \epsfbox{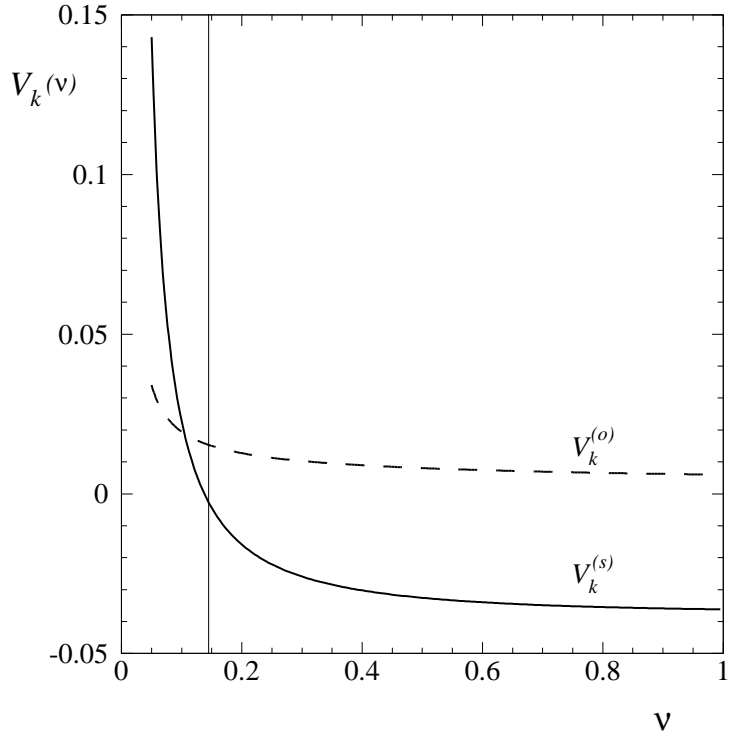}} 
{\tighten
\caption[1]{Values for the NLL running $V^{(-1)}$ potential for top quarks and
$n_f=5$ massless flavors.  The solid and dashed lines are the color singlet and
octet coefficients  for the $\pi^2/(m|{\bf k}|)$ potential. The solid vertical 
line marks the Coulombic regime where $\alpha_s(m \nu)= \nu$.}
\label{fig_vm1} }
\end{figure}
To see the effect of the running on the value of the $V^{(-1)}$
potential, consider the case of top quark production near threshold. Using
$\alpha_s(m_t)=0.108$ and Eq.~(\ref{Vkssoln}), the one loop matching value is:
\begin{eqnarray}  \label{num1}
 {\cal V}_k^{(s)}(1) = -0.0362 \,.
\end{eqnarray}
For a Coulombic system we determine the velocity $v$ by solving
$\alpha_s(mv)=v$. Using $m_t=175\,{\rm GeV}$ and the one loop running of
$\alpha_s(\mu)$ with $n_f=5$ gives $v=0.145$. The NLL color singlet and octet
coefficients for the $1/|{\bf k}|$ potential are plotted in
Fig.~\ref{fig_vm1}. At $\nu=v$ the running coupling is
\begin{eqnarray}  \label{num2}
 {\cal V}_k^{(s)}(v) = -0.0027 \,,
\end{eqnarray}
which is a substantial change from Eq.~(\ref{num1}).  In comparison, the terms
shown in Eq.~(\ref{pert1}) where the combination $\alpha_s \ln(v)$ is treated
perturbatively give ${\cal V}_k^{(s)}(v) = -0.0313$.  Thus, the summation of
logarithms for the $1/|{\bf k}|$ potential is quite important; it decreases the
coefficient by an order of magnitude.  Using the unexpanded results in
Eqs.~(\ref{VkB}) and (\ref{pertB2}) gives ${\cal V}_k^{(s)}(v) = 0.0201$, so our
resummed result is also quite different from the fixed order result in
Ref.~\cite{Brambilla}. 

\subsection{The NLL \lowercase{$t\bar t$} production current}

In terms of the running color singlet potentials the anomalous dimension for
the production current in Eq.~(\ref{pcurrent}) is\cite{LMR}:
\begin{eqnarray} \label{c1ad}
\gamma_{c_1}(\nu) = \nu {\partial  \over \partial\nu} \ln[c_1(\nu)] &=&
 -{{\cal V}_c^{(s)}(\nu)
  \over 16\pi^2} \left( { {\cal V}_c^{(s)}(\nu) \over 4 }
  +{\cal V}_2^{(s)}(\nu)+{\cal V}_r^{(s)}(\nu)
   + {\bf S}^2\: {\cal V}_s^{(s)}(\nu)  \right) +
   { {\cal V}_{k}^{(s)}(\nu) \over 2} \,. \nn\\
\end{eqnarray}
For the vector production current in Eq.~(\ref{pcurrent}) only spin-1 states
are produced, and we can set ${\bf S}^2=S(S+1)=2$. However, our analysis also
applies to the scalar production current
\begin{eqnarray} \label{spc}
  c(\nu) \: \sum_{\bf p} \: {\psip p}^\dagger {\chi^*_{\bf -p}} \,,
\end{eqnarray}
which, for example, contributes to the process $\gamma\gamma\to t\bar t$
\cite{pen}. Therefore, we will keep the factors of ${\bf S^2}$ explicit in our
results.  The boundary condition is given by the matching condition at $\nu=1$
($\mu=m$) which is known to two loops \cite{pcQCD}.  For the NLL approximation
only the one loop matching condition should be used. For the vector current
\begin{eqnarray}
  c_1(1) = 1 - \frac{2 C_F \alpha_s(m)}{\pi} \,.
\end{eqnarray}
Integrating Eq.~(\ref{c1ad}) with the running potentials in
Appendix~\ref{App_LL}, Eq.~(\ref{pv2}) and in Eq.~(\ref{Vksoln}) we find
\begin{eqnarray} \label{c1sln}
 \ln\Big[ \frac{c_1(\nu)}{c_1(1)} \Big] &=&
   a_1\:  \pi \alpha_s(m) \Big(\frac{1}{z}-1 \Big)
  + a_2\: \pi\alpha_s(m) (1-z) + a_3\: \pi \alpha_s(m) \ln(z) \nn\\
  && + a_4\: \pi\alpha_s(m) \bigg[1- z^{1-13C_A/(6\beta_0)} \bigg]
  +a_5 \: \pi\alpha_s(m) \bigg[1- z^{1-2 C_A/\beta_0}  \bigg] \nn \\
  && + a_6\: \pi\alpha_s(m) \bigg[ \frac{\pi^2}{12} -\frac12 \ln^2(2)-\ln(w)
  \ln\Big( \frac{2w}{2w-1} \Big) - \ply\Big(\frac{1}{2w}\Big) \bigg]\nn\\
  && + a_7\: \pi\alpha_s(m) \bigg[ \frac{w}{2w-1}\ln(w)-\frac12\ln(2w-1)\Big]\,,
\end{eqnarray}
where $z=\alpha_s(m\nu)/\alpha_s(m)$ and $w=\alpha_s(m\nu^2)/\alpha_s(m\nu)$.
The coefficients $a_i$ in Eq.~(\ref{c1sln}) are
\begin{eqnarray} \label{acoeffs}
  a_1 &=& { 32 C_A C_F (C_A+2 C_F) \over 3\beta_0^2 } \,,\nn\\[5pt]
  a_2 &=& { -C_F [3 \beta_0(26C_A^2+19C_A C_F-32C_F^2)+
     C_A(208 C_A^2+597 C_A C_F+716 C_F^2)]\over 78\,\beta_0^2\, C_A } \,, \nn\\
  a_3 &=& {-C_F \over 3\,\beta_0^2\, (6\beta_0-13C_A) (\beta_0-2C_A)}
      \, \Bigg\{ 2 C_F^2 (66\beta_0-179 C_A)(\beta_0 -2 C_A) \nn \\
     &&\qquad - C_A C_F \Big[ 6(49-3{\bf S^2})\beta_0^2 -
      (1126-39{\bf S^2}) \beta_0 C_A + 1067 C_A^2 \Big]\nn \\
     &&\qquad -24 C_A^2 (6\beta_0- 13 C_A)(\beta_0-2 C_A) \Bigg\}
     \,,\nn\\[5pt]
  a_4 &=& {-24 C_F^2 (3\beta_0-11 C_A)(5 C_A+8 C_F) \over 13\, C_A
     (6\beta_0-13 C_A)^2}\,,\qquad
  a_5 = {C_F^2 \Big[ (4{\bf S^2}-3) \beta_0 + (15-14{\bf S^2}) C_A \Big]
     \over 6 (\beta_0-2 C_A)^2 }\,,\nn\\[5pt]
  a_6 &=& {-16\, C_F^2 (C_A+2 C_F) \over 3\,\beta_0^2 }\,, \qquad
  a_7 = \frac{16 C_A C_F (C_A+2 C_F)}{\beta_0^2} \,.
\end{eqnarray}

A further check on our result can be made by comparing it to the
$\alpha_s^3 \ln^2(\alpha_s)$ corrections calculated by Kniehl and Penin in
Ref.~\cite{Penin}. Near threshold the $t\bar t$ cross section depends on the
product \cite{hoang}:
\begin{eqnarray}
  \sigma \propto  |c_1|^2\ \ G_C(0,0,E)\ |\psi_n^C(0)|^2 \,,
\end{eqnarray}
where  $\psi^C$ is the leading order Coulomb wavefunction, and $G_C$ is the
Coulomb Green's function. In the approach used in Ref.~\cite{Penin}, $G_C$
embodies corrections to the wavefunction at the origin, $|\psi(0)|^2\equiv
|\psi_n^C(0)|^2 [1+\Delta \psi^2(0)]$, and includes the large logarithms. They
calculate the $\alpha_s^2\ln(\alpha_s)$ and $\alpha_s^3\ln^2(\alpha_s)$ terms
and find:
\begin{eqnarray} \label{KPans}
 \Delta \psi^2(0) &=& -C_F \alpha_s^2 \ln(\alpha_s) \bigg\{ \Big[2-\frac23 {\bf
 S^2} \Big] C_F + C_A \bigg\} \nn\\
 && - \frac{C_F}{\pi} \alpha_s^3 \ln^2(\alpha_s) \bigg\{ \frac32 C_F^2 +
  \Big[ \frac{41}{12}-\frac{7}{12} {\bf S^2} \Big] C_F C_A +\frac23 C_A^2
  \bigg\} \,,
\end{eqnarray}
for the terms not involving $\beta_0$.  In our approach the large logarithms in
the cross section all appear in the running coefficient $c_1(\nu)$, so we expect
that the logarithms of Kniehl and Penin will be reproduced by
\begin{eqnarray} \label{D1}
  \Delta \psi^2(0) &=& \left| \frac{c_1(\nu)}{c_1(1)} \right|^2-1
  \,=\, 2 \ln(\alpha_s)\: \gamma_{c_1}\!(1)  +
  \ln^2(\alpha_s) \Big\{ \gamma_{c_1}'\!(1) + 2 [\gamma_{c_1}\!(1)]^2
  \Big\} + \ldots \,,
\end{eqnarray}
where we have expanded to second order in $\ln(\nu)=\ln(\alpha_s)$. The
$[\gamma_{c_1}\!(1)]^2$ term  does not contribute at order $\alpha_s^3 \ln^2
\alpha_s$, and can be dropped. The remaining terms in Eq.~(\ref{D1}) give:
\begin{eqnarray} \label{theans}
 \Delta \psi^2(0) &=& -C_F \alpha_s^2 \ln(\alpha_s) \bigg\{ \Big[2-\frac23 {\bf
 S^2} \Big] C_F + C_A \bigg\} \\
 &-& \frac{C_F}{\pi} \alpha_s^3 \ln^2(\alpha_s) \Bigg\{ \frac32 C_F^2 +
  \Big[ \frac{41}{12}-\frac{7}{12} {\bf S^2} \Big] C_F C_A +\frac23 C_A^2
  -\frac{\beta_0}{2} \bigg[ \Big( 2-\frac23 {\bf S^2} \Big) C_F+C_A \bigg] 
  \Bigg\} \,, \nn
\end{eqnarray}
which agrees exactly with the result from Ref.~\cite{Penin} in
Eq.~(\ref{KPans}) after setting $\beta_0=0$. Thus, we have shown that the logarithms of Kniehl and Penin
can indeed be associated with renormalization group logarithms. Expanding
Eq.~(\ref{c1sln}) to higher orders gives the $\alpha_s^4 \ln^3(\alpha_s)$,
$\alpha_s^5 \ln^4(\alpha_s)$, etc.\ terms.

\begin{figure}[!t]
  \centerline{\epsfxsize=10.0truecm \epsfbox{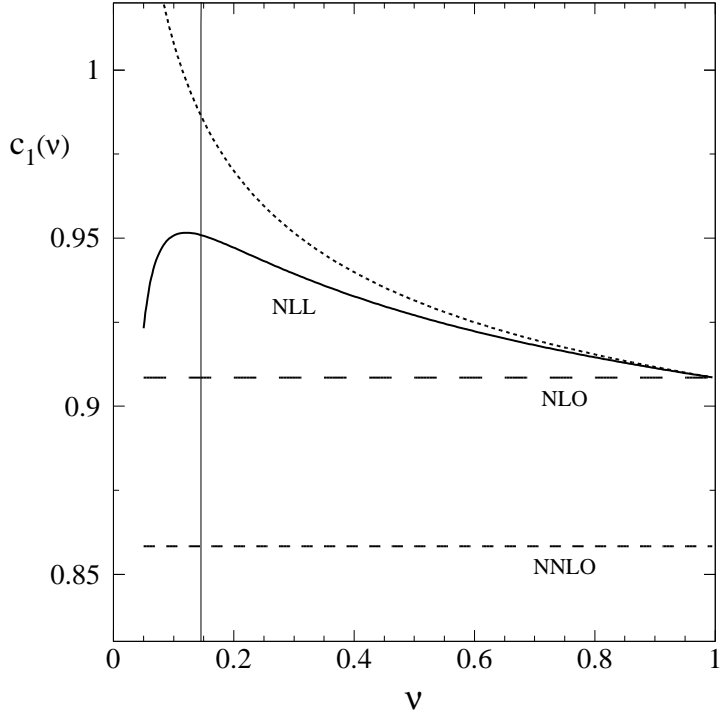}}
{\tighten \caption[1]{Values for the $t\bar t$ production current coupling
$c_1(\nu)$ for $n_f=5$ massless flavors.  At LO and LL $c_1=1$. The large and
small dashed lines show the value of $c_1$ at NLO and NNLO \cite{pcQCD}
respectively.  The solid line shows the running value at NLL order given in
Eq.~(\ref{c1sln}).  The dotted line shows the result of promoting the coupling
in the NLO result to a running coupling (the approximation used in
Ref.~\cite{BSSrun}).  The solid vertical line marks the Coulombic regime where
$\alpha_s(m \nu)= \nu$.} \label{fig_c1} }
\end{figure}

For QCD with $n_f=5$, the coefficients in Eq.~(\ref{acoeffs}) are
\begin{eqnarray}
 a_1 &=& 4.113 \,,\qquad a_2 =-2.173 \,, \qquad a_3= 4.308-0.417\,{\bf S^2} \,,
     \qquad a_4 = 5.731\,, \nn\\
 a_5 &=& 2.347-1.209\,{\bf S^2} \,,\qquad a_6= -0.914 \,,\qquad
     a_7=\phantom{-}6.170 \,.
\end{eqnarray}
The running coefficient $c_1(\nu)$ is shown in Fig.~\ref{fig_c1} for ${\bf
S^2}=2$. For comparison we have also shown by a dotted line the value of the
running coefficient obtained in Ref.~\cite{BSSrun} by an approach based on simply
taking $\alpha_s\to \alpha_s(m\nu)$ to approximate the NLL value for the
coupling. This approximation misses many of the $\alpha_s \ln(v)$ terms, and does
not provide a good estimate of the NLL result, as seen in Fig.~\ref{fig_c1}.
Some of the more important logarithms in our NLL result include the large
$\ln(m\nu^2/m)$ terms that enter through the mixing generated by ultrasoft gluon
diagrams in the potential.

We have also shown the NNLO value for $c_1$ in Fig.~\ref{fig_c1}. As pointed out
in Ref.~\cite{pcQCD} the NNLO value of $c_1(1)$ is fairly large. From
Fig.~\ref{fig_c1} we see that summing the logarithms reduced the size of the NLO
matching correction by a factor of two. It would be interesting to see if the
running induced at NNLL continues to improve the convergence of the expansion. A
consistent calculation at this order requires the three-loop anomalous dimension
of $c_1$, as well as one loop running of the $v^2$ coefficients $c_2$ and
$c_3$. This computation appears quite involved since running of the current at
this order will likely depend on the running of higher order terms in the
potential.

Several groups have analyzed the $t\bar t$ cross section predictions at NNLO
using effective field theory techniques \cite{BSSrun,tth,melnikov,yak,nag,pen}.
It should be straightforward to incorporate the renormalization group improved
current and potentials into their analysis by simply choosing a value of $\nu$
appropriate to the threshold region.  The value of $\nu$ only needs to be of
order the velocities in this region for the large logarithms to be minimized.
Additional logarithms that appear in evaluating matrix elements with the
potential will not involve large ratios of scales since they are of the form
$\ln(E/\mu_U)$ and $\ln(|{\bf k}|/\mu_S)$ where $\mu_U=m\nu^2$ and $\mu_S=m\nu$.
In a more precise analysis one might wish to use $\alpha_s(m v)C_F = v$ to
determine the value $\nu$ to use.

We would like to thank I.~Rothstein and J.~Soto for discussions.  This work was
supported in part by the Department of Energy under grant DOE-FG03-97ER40546,
and by the National Science Foundation under NYI award PHY-9457911.

\appendix

\section{LL \lowercase{values for the order \lowercase{$v^{-2}$} and
\lowercase{$v^0$} potentials}} \label{App_LL}

The leading log values for the coefficients of the $V^{(-2)}$ and $V^{(0)}$
potentials are
\cite{amis}:
\begin{eqnarray}  \label{pv2}
 {\cal V}_c^{(T)}(\nu) &=& 4 \pi \alpha_s(m)\, z \,, \\
 {\cal V}_c^{(1)}(\nu) &=& 0 \,, \nn\\
 {\cal V}_r^{(T)}(\nu) &=& 4\pi\,\alpha_s(m) z - \frac{32\pi C_A}{3\beta_0}\,
  \alpha_s(m) \Big[ 1 - z \Big]  - \frac{64\pi C_A}{3\beta_0}\,
   \alpha_s(m)\ln\Big( w \Big) \,,   \nn \\
 {\cal V}_2^{(T)}(\nu) &=& \frac{\pi\Big[ C_A (352 C_F+91 C_d-144 C_A)
    - 3\beta_0 (33 C_A+32 C_F) \Big]}{39\beta_0 C_A} \
    \alpha_s(m) \Big[z - 1 \Big] \nn\\
  &+& \frac{8\pi (3\beta_0-11C_A)(5 C_A+8 C_F) \alpha_s(m)}
    {13 C_A (6\beta_0-13 C_A)} \Big[ z^{(1-13 C_A/(6\beta_0))} - 1 \Big]
    \nn \\
  &+& \frac{ \pi (\beta_0-5 C_A)\alpha_s(m)}{(\beta_0-2 C_A)}
   \Big[ z^{(1-2C_A/\beta_0)} - 1 \Big] - \frac{8\pi (4C_F\!+\!C_d\!-\!3C_A)}
   {3\beta_0} \, \alpha_s(m) \ln\Big( w \Big)  \,, \nn \\
 {\cal V}_2^{(1)}(\nu) &=& \frac{28\pi C_1}{3\beta_0}\,
   \alpha_s(m)\Big( 1 - z \Big) +  \frac{32\pi C_1}{3\beta_0}
    \, \alpha_s(m) \ln\Big( w \Big)    \,, \nn \\
 {\cal V}_s^{(T)}(\nu) &=& \frac{2 \pi \alpha_s(m)}{(2 C_A-\beta_0) } \bigg[
   C_A + \frac{1}{3} ( 2\beta_0 - 7 C_A) \ z^{(1-2 C_A/\beta_0)} \bigg]
   + {1 \over N_c}\: \pi\, \alpha_s(m) \,, \nn\\
 {\cal V}_s^{(1)}(\nu) &=& {(N_c^2-1)\over 2N_c^2}\: \pi\, \alpha_s(m)
    \,, \nn \\
 {\cal V}_t^{(T)}(\nu) &=& -\frac{\pi\alpha_s(m)}{3}\,
    \ z^{(1-2 C_A/\beta_0)}  \,,\nn \\
 {\cal V}_\Lambda^{(T)}(\nu) &=& 2 \pi\alpha_s(m) \Big[ z - 4\
    z^{(1-C_A/\beta_0)}  \Big]  \,, \nn
\end{eqnarray}
where $z=\alpha_s(m\nu)/\alpha(m)$, $w=\alpha_s(m\nu^2)/\alpha(m\nu)$, and we
have included the $N_c$ dependent terms that come from matching the tree level
annihilation diagrams.  In the color singlet channel, ${\cal V}_t(\nu)$
and ${\cal V}_\Lambda(\nu)$ were first calculated in Ref.~\cite{chen} and
agree with Eq.~(\ref{pv2}).

\section{Potential operators that vanish on-shell} \label{App_off}

It is interesting to consider how the soft and ultrasoft anomalous dimensions
for $V^{(-1)}$ are affected when operators that vanish on-shell are included in
the potential Lagrangian, ${\cal L}_p$. These new operators can have
non-trivial anomalous dimensions.  In this Appendix we consider two examples of
how including these operators affects intermediate results, but in the end
yield the same result as the on-shell potential for observables. Equivalently,
it is shown that running the Lagrangian with the new operators from $\nu=1$ to
$\nu=v$ and then removing them by field redefinitions or operator identities
reproduces the on-shell running value of $V^{(-1)}$.

Consider the scattering $Q(p^0_1,{\bf p})+{\bar Q}(p^0_2,-{\bf p})\to
Q(p^0_3,{\bf p^\prime}) + {\bar Q}(p_4^0,-{\bf p^\prime})$. As our first
example, consider including in the $v^0$ potential in Eq.~(\ref{V0}) terms of
the form:
\begin{eqnarray} \label{V0a}
  V^{(0)} &=&
  \bigg[ {\cal V}_{\Delta 1}^{(T)}\: (T^A \otimes \bar T^A) +
  {\cal V}_{\Delta 1}^{(1)}\: (1\otimes 1) \bigg]
  { (p_3^0-p_1^0)^2 \over {\mathbf k}^4 }\ \nn\\[5pt]
  &+& \bigg[ {\cal V}_{\Delta 2}^{(T)}\: (T^A \otimes \bar T^A) +
  {\cal V}_{\Delta 2}^{(1)}\: (1\otimes 1) \bigg]
  { {(\mathbf p^{\prime 2} - p^2})^2 \over 4 m^2 {\mathbf k}^4 }\  \,.
\end{eqnarray}
We will refer to this potential as an off-shell potential, since on-shell it
vanishes by energy conservation, where $p_3^0=p_1^0$, ${\bf p'\,}^2={\bf p}^2$. 
Define $V_\Delta^{(T,1)} = V_{\Delta1}^{(T,1)}+V_{\Delta2}^{(T,1)}$.  In 
Ref.~\cite{amis2} it was shown that there is an operator identity whereby the 
time ordered product of a ${\cal V}_c$ and a ${\cal V}_\Delta$ potential gives 
${\cal V}_k$ potentials:
\begin{eqnarray} \label{kconvert}
 \begin{picture}(75,35)(-5,1)
   \put(9,16){${\cal V}_c$} \put(48,16){${\cal V}_\Delta$}
   \put(50,-1.7){$\Box$}
   \epsfxsize=2.4cm \lower9pt \hbox{\epsfbox{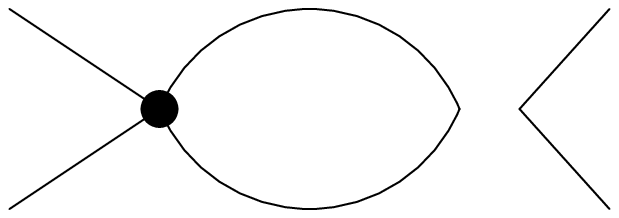}}
 \end{picture} +
 \begin{picture}(75,35)(-5,1)
   \put(9,16){${\cal V}_\Delta$} \put(48,16){${\cal V}_c$}
   \put(9,-1.7){$\Box$}
   \epsfxsize=2.4cm \lower9pt \hbox{\epsfbox{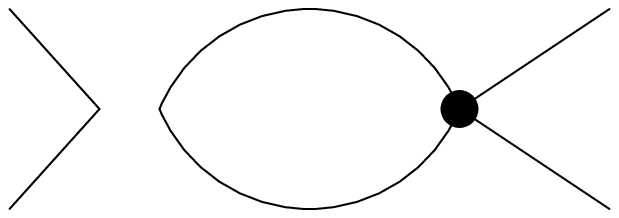}}
 \end{picture}
   &=& \frac{i {\cal V}_c^{(T)} {\cal V}_\Delta^{(T)} }{32 m k}\ T^A T^B \otimes
      \bar T^A \bar T^B +\frac{i {\cal V}_c^{(T)} {\cal V}_\Delta^{(1)} }
     {32 m k}\ T^A \otimes \bar T^A +\ldots \,, \nn\\
\end{eqnarray}
where the ellipses denote additional terms which vanish on-shell. The identity
in Eq.~(\ref{kconvert}) allows us to remove the potential in Eq.~(\ref{V0a}) at
an arbitrary velocity scale $\nu$ in favor of a purely on-shell potential.
Transforming ${\cal V}_\Delta^{(T,1)}(\nu)\to 0$ gives
\begin{eqnarray}  \label{trnsfm}
 {\cal V}_k^{(T)}(\nu) & \to & {\cal V}_k^{(T)}(\nu) + \frac{1}{32\pi^2}\:
   {\cal V}_c^{(T)}(\nu) \bigg[ -{\cal V}_\Delta^{(1)}(\nu) +\frac14 (C_A+C_d)
   {\cal V}_\Delta^{(T)}(\nu) \Big] \,, \nn \\*
 {\cal V}_k^{(1)}(\nu) & \to & {\cal V}_k^{(1)}(\nu) - \frac{1}{32\pi^2}\:
   {\cal V}_c^{(T)}(\nu)\: C_1 \: {\cal V}_\Delta^{(T)}(\nu)  \,. 
\end{eqnarray}
In a situation where ${\cal V}_\Delta^{(1)}(\nu)= 0$ transforming ${\cal
V}_\Delta^{(T)}(\nu)\to 0$ induces a color singlet $1/|{\bf k}|$ potential of
the form
\begin{eqnarray} \label{trnsfm2}
 {\cal V}_k^{(s)}(\nu) & \to & {\cal V}_k^{(s)}(\nu) - \frac{C_F^2}{32\pi^2}\:
   {\cal V}_c^{(T)}(\nu) {\cal V}_\Delta^{(T)}(\nu) \,. 
\end{eqnarray}
(Transformations between order $1/v$ and $v^0$ potentials were also considered
in Refs.~\cite{hoang,melnikov,mm1bram}.) 

To make the implications of Eq.~(\ref{trnsfm}) more clear, we will consider an
example and show that including the $V_\Delta^{(T,1)}$ potential will not effect
predictions for observables when running below $m$. Consider matching offshell
in Feynman gauge where the matching coefficients at $\nu=1$ are $V_\Delta^{(T)}
= 4\pi\alpha_s(m)$ and $V_\Delta^{(1)}=0$, and the one loop anomalous dimensions
for the potentials in Eq.~(\ref{V0a}) are~\cite{amis2}:
\begin{eqnarray} \label{adVd}
  \nu {\partial \over \partial\nu} {\cal V}_{\Delta}^{(T)} &=& -2 \beta_0
  \alpha_s(m \nu)^2 \,, \nn\\
  \nu {\partial \over \partial\nu} {\cal V}_{\Delta}^{(1)} &=& 0
  \,. \label{convanom}
\end{eqnarray}
This anomalous dimension has contributions from the one loop graphs with two
soft gluons, ghosts or quarks where: the two factors of ${(\mathbf p^{\prime 2}
- p^2})$ are from the soft vertices, or these factors are from two insertions
on the soft propagators, or where one factor is from the propagator and one
from the vertex (for the ghost loop). The solution of Eq.~(\ref{adVd}) is
$V_\Delta^{(T)}(\nu)=4\pi\alpha_s(m\nu)$ and $V_\Delta^{(1)}(\nu)=0$. The
one loop counterterm which generates this running affects our soft anomalous
dimension computation since now there are counterterms of the form
\begin{eqnarray}
  { ({\bf p'\,^2-p^2})^2 \over m^2 (\bf p'-p)^4\ \epsilon }  \,.
\end{eqnarray}
These counterterm give rise to one loop diagrams which exactly cancel the
divergences in Eq.~(\ref{Fig3a}) for the two loop graph in Fig.~\ref{fig_s}a.
Recalling that the one loop matching values for the $1/|{\bf k}|$ potentials are
also different~\cite{amis2}, we find that the running ${\cal V}_k^{(T,1)}(\nu)$
coefficients in Eq.~(\ref{Vksoln}) become
\begin{eqnarray} \label{Vksoln2}
  {\cal V}_k^{(T)}(\nu)  &=& \frac{(3 C_A-C_d)}{4}\: \alpha_s^2(m)
   + \left[\frac{(3C_A-C_d)}{4}+\frac{8\,C_A(C_A+C_d)}{3\beta_0}\right]
   ( z^2 - 1 )\: \alpha_s^2(m)  \\
   &&+ \frac{8 C_A(C_A+C_d)}{\beta_0} \Big[ z-1-2\ln(w)\Big]\:\alpha_s^2(m)
   \,, \nn\\
  {\cal V}_k^{(1)}(\nu) &=& {C_1} \: \alpha_s^2(m)
   +\left[ {C_1} -\frac{16\,C_A C_1}{\beta_0}\right] (z^2 - 1)\:
   \alpha_s^2(m) 
   - \frac{48 C_A C_1}{\beta_0}\: \alpha_s^2(m) \Big[ z-1-2\ln(w) \Big] \,,\nn
\end{eqnarray}
This Feynman gauge off-shell potential is consistent with the transformation in
Eq.~(\ref{trnsfm}) which transforms the off-shell case, $V_\Delta^{(T)} =
4\pi\alpha_s(m\nu)$ and Eq.~(\ref{Vksoln2}), into the on-shell case,
$V_\Delta^{(T,1)}=0$ and Eq.~(\ref{Vksoln}). The comparison in
section~\ref{nll1} with Ref.~\cite{Brambilla} shows that our on-shell analysis is
also consistent with an off-shell Coulomb gauge potential.

With the off-shell potential in Eq.~(\ref{V0a}) there are two new
graphs which contribute to the anomalous dimension for the production current:
\begin{eqnarray}
\begin{picture}(180,30)(160,1)
  \put(186,1.7){$\Box$} \put(335,1.7){$\Box$}
  \put(137,20){${\cal V}_c$}      \put(180,20){${\cal V}_\Delta$}
  \put(332,20){${\cal V}_\Delta$} \put(380,20){${\cal V}_c$}
  \centerline{
   \epsfxsize=5cm \lower9pt \hbox{\epsfbox{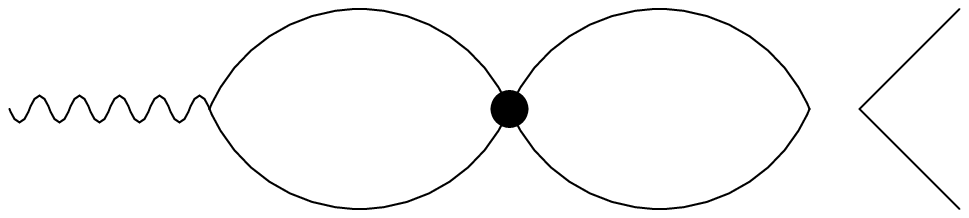}} \qquad\qquad
   \epsfxsize=5cm \lower9pt \hbox{\epsfbox{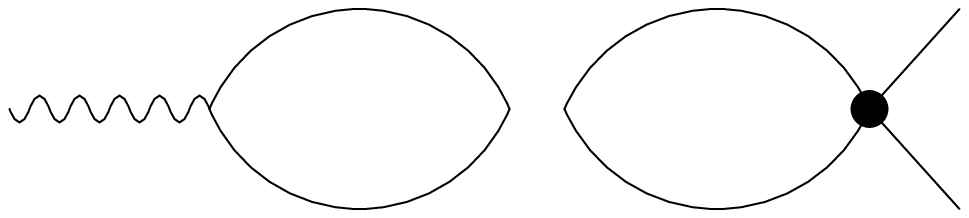}}
   } $\!\!\!\!\!\!\!\!\!\!\!\!\!\!\!\! \,,$
\end{picture}
\end{eqnarray}
and the anomalous dimension is therefore
\begin{eqnarray} \label{pc2}
 \nu {\partial \over \partial\nu} \ln[c_1(\nu)] &=& {-{\cal V}_c^{(s)}(\nu)
  \over 16\pi^2} \left( { {\cal V}_c^{(s)}(\nu) \over 4 }
  +{\cal V}_2^{(s)}(\nu)+{\cal V}_r^{(s)}(\nu)
   + {\bf S^2} {\cal V}_s^{(s)}(\nu)  \right) +
   { {\cal V}_{k}^{(s)}(\nu) \over 2}  - \frac{ {\cal V}_c^{(s)}\:
   {\cal V}_\Delta^{(s)} }{64 \pi^2} \,, \nn\\
\end{eqnarray}
where  ${\cal V}^{(s)}_\Delta(\nu)={\cal V}^{(1)}_\Delta(\nu) - C_F {\cal
V}^{(T)}_\Delta(\nu)$.  Since the last two terms in Eq.~(\ref{pc2}) are
invariant under the transformation in Eq.~(\ref{trnsfm2}) the off-shell
potential gives the same prediction for the running of the production current.
In calculating observables the operator identity in Eq.~(\ref{kconvert})
guarantees that the time ordered product of the running ${\cal V}_\Delta^{(T)}$
and ${\cal V}_c^{(T)}$ produces the same effect as the soft contribution to the
running of the on-shell ${\cal V}_k^{(T)}$ in Eq.~(\ref{Fig3a}).

As our second example, consider including an operator which would vanish by
the lowest order free equations of motion:
\begin{eqnarray} \label{Loff}
  {\cal L}_{\rm eom} = -\sum_{\bf p,p'} \frac{{\cal V}_F }{({\mathbf p'-p})^2}
  \bigg[ \psip{p'}^\dagger \, T^A \Big( i\partial_0-\frac{\bf p^2}{2m} \Big)
  \psip p \bigg] \chip{-p'}^\dagger \, \bar T^A\, \chip{-p} \ \  + \ \
  \chi \leftrightarrow \psi  \,.
\end{eqnarray}
The only feature of the operator in Eq.~(\ref{Loff}) that is different from
typical operators that vanish by the equations of motion (for example, in
HQET\cite{run}) is its non-local nature relative to the scales ${\bf p} \sim
{\bf p'} \sim m v$. At one loop there are ultrasoft gluon graphs which mix
into the operator in Eq.~(\ref{Loff}), for instance from the diagrams:
\begin{eqnarray} \label{off2}
\begin{picture}(160,40)(160,1)
  \put(70,20.7){$a)$} \put(250,20.7){$b)$}
    \put(321,-9){${\cal V}_c$}
  \put(338,18.5){$\mbox{\large $\mathbf{\times}$}$}
   \centerline{
   \epsfxsize=3cm \lower9pt \hbox{\epsfbox{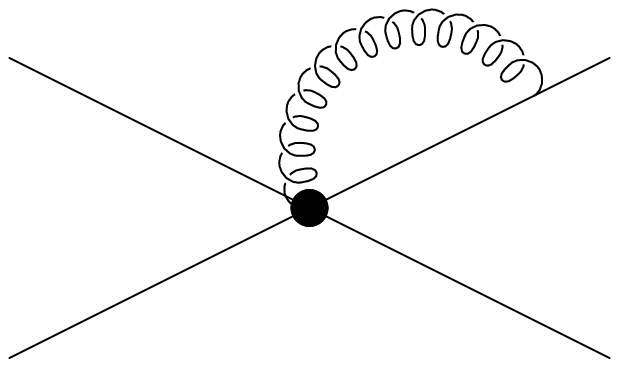}}
    \qquad\qquad\qquad\qquad
   \epsfxsize=3cm \lower9pt \hbox{\epsfbox{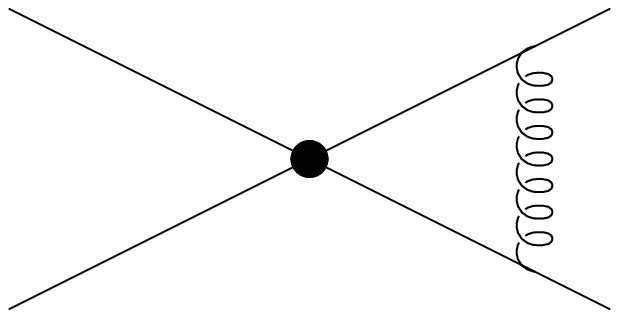}}
   } $ \!\!\!\!\!\!\!\!\!\!\!\!\!\!\!\!\!\!\!\!\!\!\!\!\!\!\!\!\!\!\! \,.$
\end{picture}
\end{eqnarray}
In the first graph the vertex is the first term in Eq.~(\ref{Lpu}) and the
ultrasoft gluon is ${\bf A}$, while in the second diagram the cross is a
insertion of the ultrasoft $\nabla^2/m$ operator in Eq.~(\ref{Lu}) and the gluon
is an $A^0$. In Feynman gauge both graphs produce a divergence of the form
\begin{eqnarray}
  \frac{ (E-{\bf p^2}/m) }{ \epsilon\: ({\bf p'-p})^2 } \,,
\end{eqnarray}
which vanishes by the equations of motion, but off-shell renormalizes the
coefficient of the operator in Eq.~(\ref{Loff}). To see how this
effects the calculation of our ultrasoft anomalous dimension, consider
Coulomb gauge, where only Eq.~(\ref{off2}a) is non-zero.  In this case there
is a one loop counterterm diagram which exactly cancels the divergence in
Fig.~(\ref{Cusoft}b). After running down to the low scale the value of
${\cal V}_k^{(T,1)}(\nu)$ are different, but our potential also includes the
operator in Eq.~(\ref{Loff}) with coefficient ${\cal V}_F(\nu)$. If we wish
to remove the operator in Eq.~(\ref{Loff}) we can do so with a field
redefinition
\begin{eqnarray}
  \psip{p}^\dagger \to \psip{p}^\dagger + \sum_{\bf p''} \frac{ {\cal V}_F }
    {({\bf p'-p})^2} \psip{p''}^\dagger T^A \chip{-p''}^\dagger\, \bar T^A
    \chip{-p} \,,
\end{eqnarray}
which induces a term from the $\psip{p}^\dagger [ i\partial_0 -
{\bf p^2}/(2m) ] \psip{p}$ Lagrangian which cancels Eq.~(\ref{Loff}). This
field redefinition also gives other new contributions. For us the important
point is that the Coulomb potential induces a six-quark operator of the form:
\begin{eqnarray} \label{opw}
  -\sum_{\bf p,p'} \sum_{\bf p''} \psip{p''}^\dagger \chip{-p''}^\dagger
   \chip{-p'} \psip{p} \chip{-p'}^\dagger \chip{-p}
  \frac{ {\cal V}_F {\cal V}_c} { ({\bf p''-p'})^2\: ({\bf p'-p})^2 } \,.
\end{eqnarray}
(The contraction of color indices and factors of $T^A$ have been suppressed.)
Usually a six quark operator could not possibly effect the running of a four
quark operator such as the $1/|{\bf k}|$ potential. However, because of the
momentum dependence in the denominator of Eq.~(\ref{opw}) it induces a non-zero
tadpole diagram where the fields $\chip{-p'}$ and $\chip{-p'}^\dagger$ are
contracted. This tadpole graph produces a $1/|{\bf k}|$ and makes up for the
running in ${\cal V}_k$ that was removed when the contribution from
Eq.~(\ref{Cusoft}b) was canceled by a ${\cal V}_F$ counterterm.

{\tighten

} 

\end{document}